\documentclass[preprint,12pt]{elsarticle}
\usepackage[utf8]{inputenc}
\usepackage{appendix}
\usepackage{bm}
\usepackage{epsf,epsfig,float,latexsym}
\usepackage{amssymb}
\usepackage{graphicx}
\usepackage{amsmath}
\usepackage{amssymb}
\usepackage{rotating} 
\usepackage{booktabs} 
\usepackage{url}
\usepackage{xcolor}
\usepackage{dsfont}

\newcommand{\beq}{\begin{equation}}
\newcommand{\eeq}[1]{\label{#1}\end{equation}}
\newcommand{\beqa}{\begin{eqnarray}}
\newcommand{\eeqa}[1]{\label{#1}\end{eqnarray}}
\newcommand{\eeqan}{\end{eqnarray}}

\journal{Nuclear Physics A}

\date{\today}

\begin{document}

\begin{frontmatter}

\title{SU(3) flavor symmetry considerations for the $\bar{K}N$ coupled channels system}

\author{P.~C.~Bruns}
\author{A.~Ciepl\'{y} \corref{correspondence}}
\cortext[correspondence]{Corresponding author}
\ead{cieply@ujf.cas.cz}

\address{Nuclear Physics Institute of the Czech Academy of Sciences, 250 68 \v{R}e\v{z}, Czechia}

\begin{abstract}
We study the impact of SU(3) flavor symmetry breaking on the properties of dynamically 
generated $\Lambda^*$ states within an effective separable potential model describing 
the coupled channels $\bar{K}N$ system. The model is based on the chiral meson-baryon Lagrangian 
at next-to-leading order, and constitutes an improvement over a previous model developed 
by our group, with its applicability being extended to higher energies covering the $\Lambda(1670)$ 
resonance region. It is demonstrated that the ratios of channel couplings to the resonant states 
can vary dramatically when the flavor breaking is gradually switched off, tracing a path 
to the restored SU(3) symmetry. We conclude that the couplings determined from physical 
observables cannot be used to reliably relate a given resonance to a specific flavor multiplet.
\end{abstract}


\end{frontmatter}

%%%%%%%%%%%%%%%%%%%%%%%%%%%%%%%%%%%%%%%%%%%%%%%%%%%%%%%%%%%%%%%%%%%%%%%%%%%%%%%%%%%%%%%%%%%%%%%
\section{Introduction}
\label{sec:intro}
%%%%%%%%%%%%%%%%%%%%%%%%%%%%%%%%%%%%%%%%%%%%%%%%%%%%%%%%%%%%%%%%%%%%%%%%%%%%%%%%%%%%%%%%%%%%%%%

In the limit of exact SU(3) flavor symmetry ($m_{u}=m_{d}=m_{s}$), and when only the strong 
interaction is relevant, all hadrons can be classified according to the flavor multiplet 
they belong to \cite{Gell-Mann:1961omu}. Considering the real world, where flavor symmetry is broken, 
it appears that this classification is still very useful for a basic understanding of the structure 
of hadrons, even though it is only an approximate one. In a framework employing chiral effective 
Lagrangians \cite{Weinberg:1978kz, Gasser:1984gg, Jenkins:1990jv, Krause:1990xc, Bernard:1995dp} 
(as well as in Lattice QCD), one can even tune the flavor breaking terms away from their physical 
values, and establish a connection to a hypothetical flavor-symmetric world, where the flavor 
classification strictly applies. In the present work, we study the impact of such a procedure 
on $\Lambda^*$ resonances (isoscalar baryons with strangeness $S=-1$) that are dynamically 
generated in a coupled-channel framework based on effective potentials derived from a chiral Lagrangian 
\cite{Kaiser:1995eg,Oset:1997it,Oller:2000fj,Oset:2001cn,Jido:2003cb}.
An earlier study in this direction can be found in Sec.~3 of Ref.~\cite{Jido:2003cb}, where only 
a leading order (LO) Weinberg-Tomozawa (WT) interaction kernel was employed. There, an SU(3) flavor 
singlet and two degenerate octet states were generated in the flavor-symmetric limit.
Here we aim at extending such an analysis by the inclusion of Born and next-to-leading-order (NLO) 
contact terms in the interaction kernel, which are indispensable to obtain a satisfying description 
of the data from the kaon-nucleon threshold up to the region of the $\Lambda(1670)$ resonance. 

The early analyses \cite{Kaiser:1995eg,Oset:1997it} based on chirally motivated coupled-channels 
$\bar{K}N$ interactions concentrated on reproducing the $K^{-}p$ total cross sections at low kaon 
momenta and the threshold branching ratios, fitting the model parameters to relatively old bubble 
chamber data. The later measurements of kaonic hydrogen characteristics by the DEAR \cite{DEAR:2005fdl} 
and SIDDHARTA \cite{Bazzi:2011zj} collaborations put more stringent constraints on the theoretical 
models, though the focus remained on the energy region dominated by the $\Lambda(1405)$ resonance 
\cite{Ikeda:2012au, Mai:2012dt}. 
This is also the case with the Prague model \cite{Cieply:2011nq}, which we use as a guideline  
for our current analysis and for testing the model dependence of the presented results. The particular 
model used in our study is based on the same framework we adopted for treating the $\eta N - \eta'N$ 
interactions and aims at covering a broader energy region, from the $\bar{K}N$ threshold 
up to the $\Lambda(1670)$ domain. This also follows a recent trend of extending the available 
chirally motivated $\bar{K}N$ models to energies much higher than those dominated by the $\Lambda(1405)$, 
see e.g.~\cite{Guo:2012vv, Feijoo:2018den, Feijoo:2021zau}.

A novel feature of the present work lies in an emphasis on the behavior of ratios of couplings 
of the resonances to various meson-baryon (MB) states. These ratios are tightly constrained in the fully 
flavor-symmetric case, and we assess the applicability of the SU(3)-relations in the presence of realistic 
flavor-symmetry breaking. To achieve this task we utilize a new coupled-channel $\bar{K}N$ model 
described in some detail in Sec.~\ref{sec:model}. In Sec.~\ref{sec:SU3}, we discuss the SU(3) 
symmetry constraints on the resonance couplings. There, we restrict ourselves to some symmetry relations 
that will be needed for the discussion of our results in Sec.~\ref{sec:results}. Then, we present 
the outcome of our updated fits, and discuss the behavior of the resulting resonance pole positions 
and couplings when the flavor-breaking is continuously switched off. This allows us to relate 
our dynamically generated resonances to pertinent flavor multiplets. The conclusions we draw from 
our findings are provided in Sec.~\ref{sec:summary}.

%%%%%%%%%%%%%%%%%%%%%%%%%%%%%%%%%%%%%%%%%%%%%%%%%%%%%%%%%%%%%%%%%%%%%%%%%%%%%%%%%%%%%%%%%%%%%%%
\section{Coupled channels $\bar{K}N$ model description}
\label{sec:model}
%%%%%%%%%%%%%%%%%%%%%%%%%%%%%%%%%%%%%%%%%%%%%%%%%%%%%%%%%%%%%%%%%%%%%%%%%%%%%%%%%%%%%%%%%%%%%%%

The notation and conventions follow closely those detailed in our previous work 
on the $\eta N - \eta'N$ coupled channels system \cite{Bruns:2019fwi}. Here we just 
disregard the $\eta - \eta'$ mixing, assuming the $\eta$ meson involved in channels coupled 
to the $\bar{K}N$ system is represented by a pure octet state. The tree-level contributions 
derived from a manifestly Lorentz invariant chiral Lagrangian are written in a form
\beq
v_{0+}(s) 
= \frac{\sqrt{E+m}}{F_{\Phi}}\left(\frac{C(s)}{8\pi\sqrt{s}}\right)\frac{\sqrt{E+m}}{F_{\Phi}} 
\eeq{eq:poten}
where we employ a convenient channel-matrix notation. The diagonal matrices $E$, $m$ 
and $F_{\Phi}$ are assembled from the baryon center-of-mass (c.m.) energies, baryon masses 
and meson decay constants of the respective channels. The channel matrix $C(s)$ 
contains all the details specific to the effective vertices and the various elastic 
and inelastic MB reactions. In some more detail,
\begin{eqnarray}
  C(s) \!&=&\! \frac{1}{4}\lbrace(\sqrt{s}-m),\,C^{WT}\rbrace -  C^{NLO}(s) \nonumber \\
       \!&-&\! \frac{(\sqrt{s}-m)C^{BD}(\sqrt{s}-m)}{\sqrt{s}+m_{c}} - C^{BC}(s)  \label{eq:f0pPOT2} \; , \\
  C^{NLO}(s) \!&=&\! 2M_{\pi}^2 \,C^{\pi} + 2M_{K}^2 \,C^{K} - 2q^{0}(s)\:C^{d}\:q^{0}(s) \; . \nonumber 
\end{eqnarray}
The $q^{0}$ represents the diagonal channel matrix $q^{0}(s)=(s-m^{2}+M^{2})/(2\sqrt{s})\,$,
featuring the meson c.m.~energies in the respective MB channels, while $s$ is the usual
Mandelstam variable given by the square of the two-body c.m.~energy and $M$ is the meson mass matrix. The channel matrices $C^{WT}$,
$C^{NLO}(s)$, $C^{BD}$ and $C^{BC}(s)$ contain the couplings derived from the WT term,
the contact NLO terms, and the Born direct and crossed terms, respectively. Their exact form 
appropriate for the $\bar{K}N$ system can be found e.g.~in \cite{Borasoy:2005ie}. We also note 
that when dealing with the crossed Born term $C^{BC}(s)$ we adopt the prescription provided 
in Appendix D in \cite{Bruns:2019fwi} but keep only the term linear in energy, omitting 
the last quadratic contribution given in Eq.~(D.10) there. This approximation was introduced 
to tame unphysical divergences at subthreshold energies appearing when an exact BC-term contribution 
is projected onto the s-wave. We checked that the approximated formula works quite well 
and reproduces the exact form in a broad interval of energies.

Following Refs.~\cite{Cieply:2011nq} and \cite{Bruns:2019fwi} the s-wave 
scattering amplitude for isospins $I=0$ and $I=1$ is obtained as an algebraic solution 
of the Lippmann-Schwinger equation with the interaction kernel constructed from 
the tree-level amplitudes, Eq.~(\ref{eq:poten}), by sandwiching them with the Yamaguchi form-factors 
\beq
g_{jb}(p) = \left(1 + p^2/\alpha^{2}_{jb} \right)^{-1}\;,
\eeq{eq:YFF}
that introduce channel dependent inverse ranges $\alpha_{jb}$, where $p$ stands for the modulus 
of the momentum of meson $j$ in the c.m.~frame. At the same time, the parameters $\alpha_{jb}$ 
serve as soft cutoffs to regularize the MB loop functions $G(s)$. 
The scattering amplitude is then obtained in a separable form
\beq
f_{0+}^{I}(s) = g(s) \,\lbrack 1-v_{0+}^{I}(s)\,G(s)\rbrack^{-1} \,v_{0+}^{I}(s) \,g(s).
\eeq{eq:f0pMOD}
Here $g(s)$ is the Yamaguchi form factor with $p$ replaced by $q(s)$, the modulus of the on-shell 
MB c.m.~three-momentum in the respective channel. The diagonal matrix $G(s)$ 
is composed from MB loop functions and can be evaluated analytically, see Eq.~(8) in \cite{Bruns:2019fwi}.
One can easily check that $f_{0+}^{I}$ 
satisfies  partial-wave unitarity in the considered two-body channels space.

The low energy constants (LECs) that provide couplings for the terms included in the chiral Lagrangian 
and the inverse ranges introduced to regularize the loop function integrals need to be determined 
in fits to experimental data. When dealing with $\bar{K}N$ interactions at low energies one standardly 
includes the experimental data on total cross sections for reactions going from the initial $K^{-}p$ 
state to all channels that are open at the threshold (or slightly above it in case of $\bar{K}^{0}n$) 
\cite{Csejthey-Barth:1965izu, Sakitt:1965kh, Kim:1965zzd, Mast:1975pv, Bangerter:1980px, Ciborowski:1982et}. 
These are complemented by the data on three threshold branching ratios $\gamma$, $R_c$ and $R_n$ 
\cite{Nowak:1978au, Tovee:1971ga}, and on the measurement of the characteristics of the kaonic hydrogen 
atom, the 1s level energy shift $\Delta E_N(1s)$ and absorption width $\Gamma (1s)$ caused 
by strong interaction \cite{Bazzi:2011zj}. As we aim at performing a more complete analysis 
of dynamically generated $\Lambda^{*}$ resonances, it is desirable to reproduce well not only 
the $\Lambda(1405)$ but cover the energies where the $\Lambda(1670)$ resonance manifests itself too. 
For this reason we also include in our fits the data on the total cross sections for reactions 
that go to the $\eta \Lambda$ and $\eta \Sigma$ channels \cite{Baxter:1973ggf, Jones:1974si, CrystalBall:2001uhc}.
Since our $\bar{K}N$ model contains a relatively large number of parameters, it is essential 
to fix some of them to already established values. By doing so we reduce the number 
of degrees of freedom which in turn provides a better control over the fitting procedure. 
First of all it seems natural to adopt the following:
\begin{itemize}
\item Meson decay constants $F_{\pi}=92.3$ MeV, $F_{K}=1.193\cdot F_{\pi}=110.1$ MeV, 
$F_{\eta}=1.28\cdot F_{\pi}=118.3$ MeV complying with the PDG \cite{Rosner:2018pdg} 
and the lattice results \cite{Aoki:2019cca}.
\item The MB axial couplings $F=0.46$ and $D=0.80$ as extracted in analyses of hyperon decays \cite{Ratcliffe:1998su}.
\item $b_D = 0.1$ GeV$^{-1}$, about the average value from various fits and estimates available in the literature. 
The $b_0$ and $b_F$ are left to be determined in the fits as they appear less stable due to renormalization 
caused by loop function contributions \cite{Bruns:2019fwi}.
\end{itemize}
This setting leaves us with 12 parameters to be determined in the fits: 6 inverse ranges $\alpha_{jb}$, 
4 $d$-couplings, $b_0$ and $b_F$. 

The fit was performed using the MINUIT routine from the CERNLIB library to minimize the $\chi^{2}$ 
per degree of freedom defined as
\beq
\chi^{2}/dof = \frac{\sum_{i}N_{i}}{N_{obs}(\sum_{i}N_{i}-N_{par})} \sum_{i}\frac{\chi^{2}_{i}}{N_{i}} \; 
\eeq{eq:chi2}
where $N_{par}$ is the number of fitted parameters, $N_{obs}$ is a number of observables, $N_{i}$ 
is the number of data points for an $i$-th observable, and $\chi^{2}_{i}$ stands for the total $\chi^{2}$ 
computed for the observable. Eq.~(\ref{eq:chi2}) guarantees an equal weight of the fitted data from 
various processes (i.e. for different observables). We have achieved $\chi^{2}/dof = 1.31$, and the fitted 
parameters are provided in Table \ref{tab:newfit} aside the values adopted in a previous Prague fit tagged 
as the NLO30 model in \cite{Cieply:2011nq}. In the latter, the fitting procedure was different 
and only 7 parameters were fitted to the data that did not include the cross sections to the $\eta \Lambda$ 
and $\eta \Sigma$ channels that open at higher energies. Considering the differences between the current 
and NLO30 model fit procedures it is not surprising that the two parameter sets differ significantly 
with respect to some parameter values. In particular it looks that the inverse ranges $\alpha_{jb}$ 
for channels opening at higher energies were underestimated in the NLO30 model.

%----------------------------------------------------------------------------------------------
\begin{table}[htb]
\caption{The inverse ranges (in MeV) and NLO couplings (in GeV$^{-1}$) obtained in fits of low energy $K^{-}p$ data. 
The parameters used by the previous NLO30 Prague model \cite{Cieply:2011nq} are shown for comparison too. 
The values marked with an asterisk were kept fixed in the respective fits.}
\begin{center}
\begin{tabular}{c|cc}
 parameter               & current fit & NLO30 fit \\ \midrule
 $\alpha_{\pi \Lambda}$  &  400.0 &  297.0 \\
 $\alpha_{\pi \Sigma}$   &  509.2 &  490.6 \\
 $\alpha_{\bar{K}N}$     &  752.0 &  699.7 \\
 $\alpha_{\eta \Lambda}$ &  979.7 &  700.0$^{\ast}$ \\
 $\alpha_{\eta \Sigma}$  &  797.2 &  700.0$^{\ast}$ \\
 $\alpha_{K \Xi}$        & 1078.5 &  700.0$^{\ast}$ \\
 $d_1$                   & -0.119 & -0.368 \\
 $d_2$                   &  0.074 &  0.048 \\
 $d_3$                   &  0.096 &  0.088 \\
 $d_4$                   &  0.556 & -0.358 \\
 $b_0$                   &  0.525 & -0.321$^{\ast}$ \\
 $b_D$                   &  0.100$^{\ast}$ &  0.064$^{\ast}$ \\
 $b_F$                   & -0.077 & -0.209$^{\ast}$ 
\end{tabular}    
\end{center}
\label{tab:newfit}
\end{table}
%----------------------------------------------------------------------------------------------

\begin{figure}[htb!]    
\centering
\includegraphics[width=0.42\textwidth]{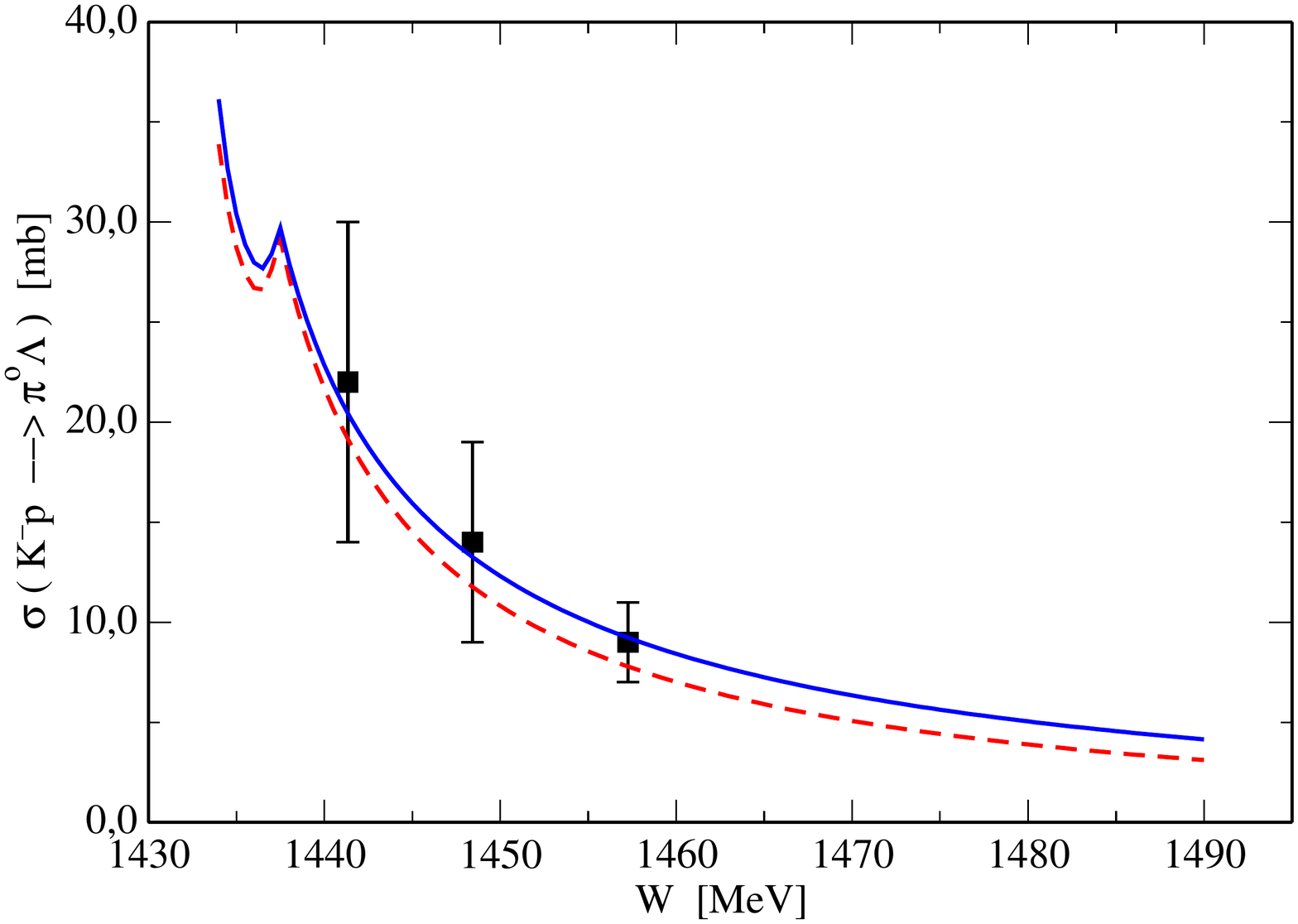}
\includegraphics[width=0.42\textwidth]{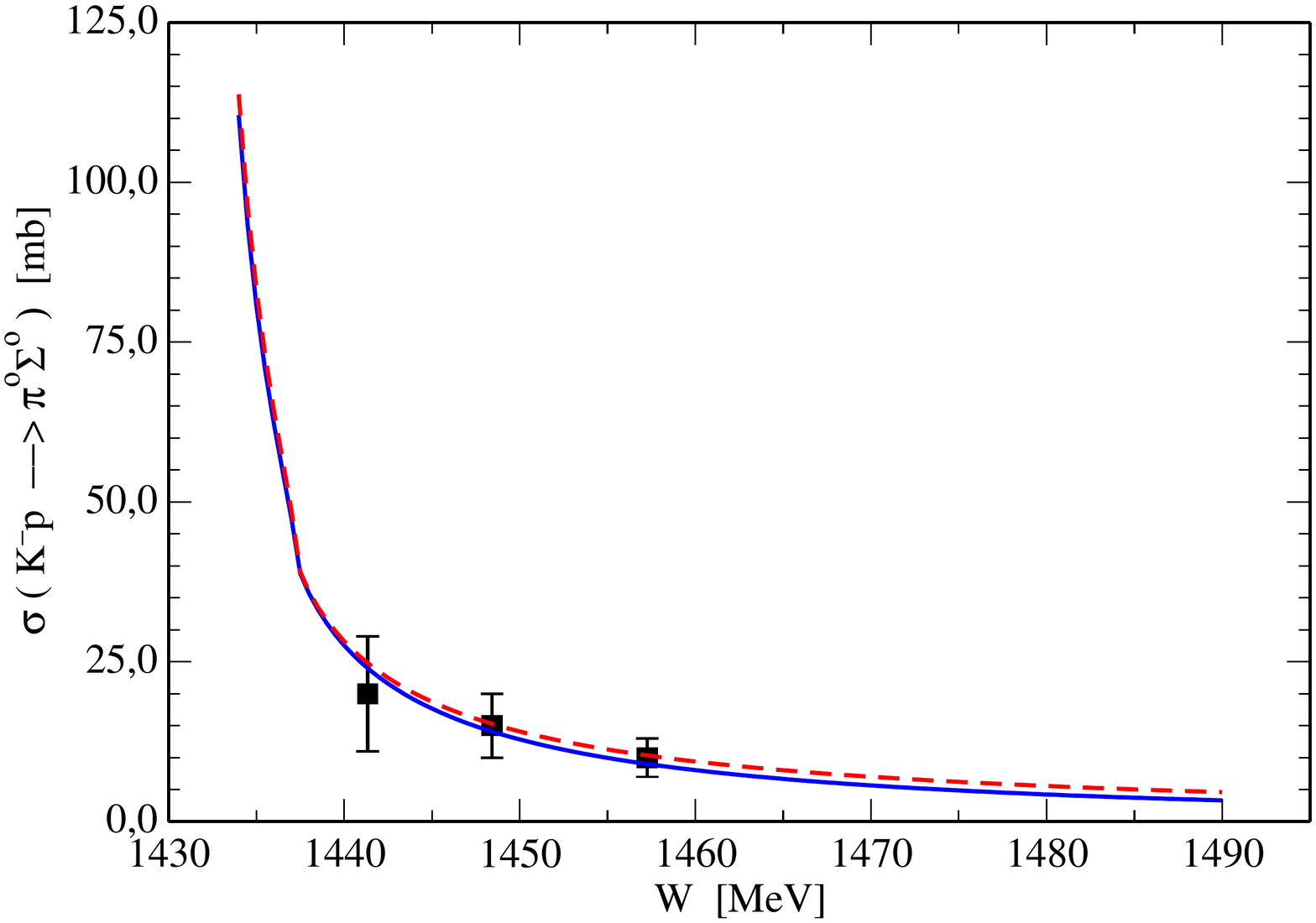} \\[-3mm]
\includegraphics[width=0.42\textwidth]{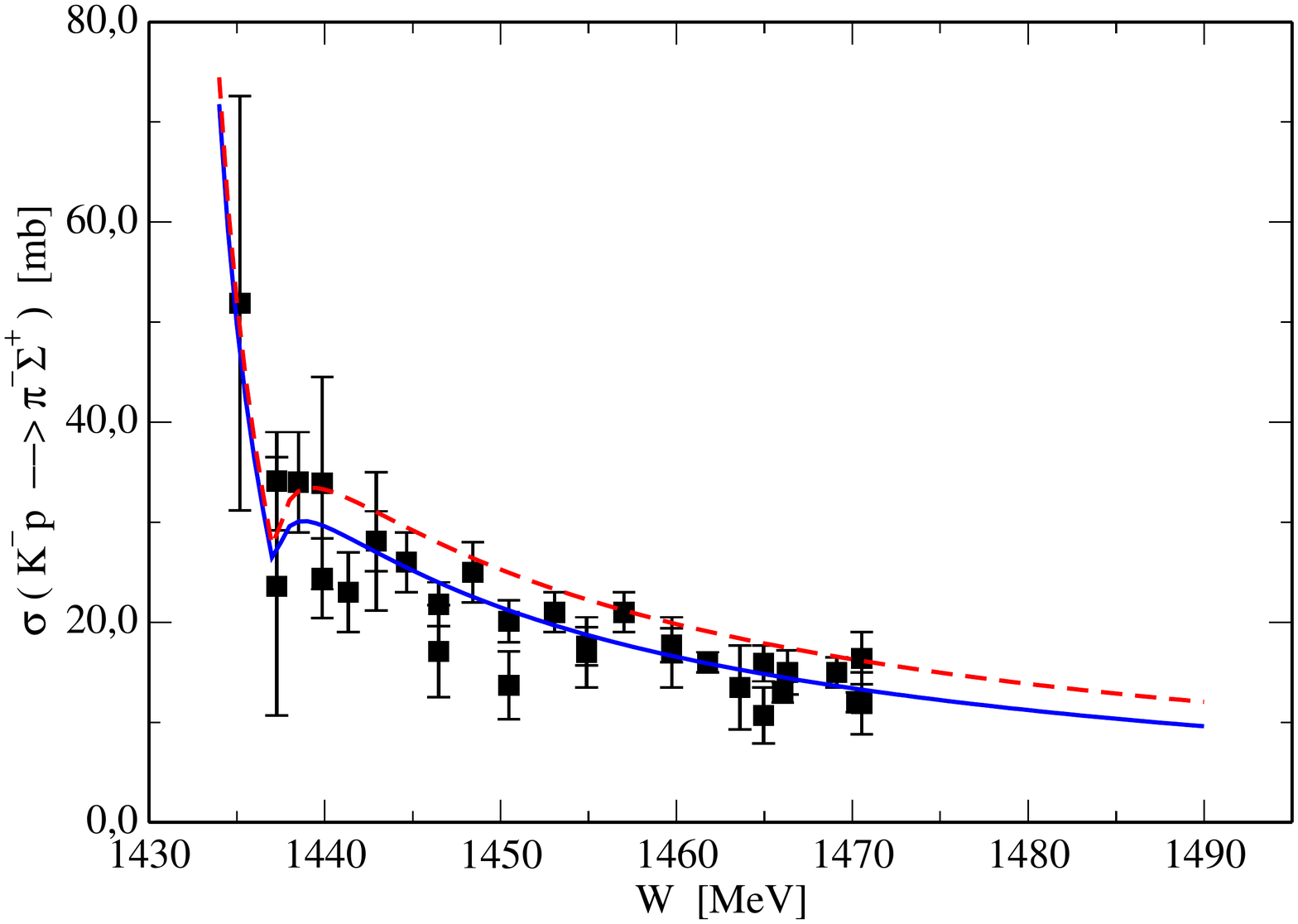}
\includegraphics[width=0.42\textwidth]{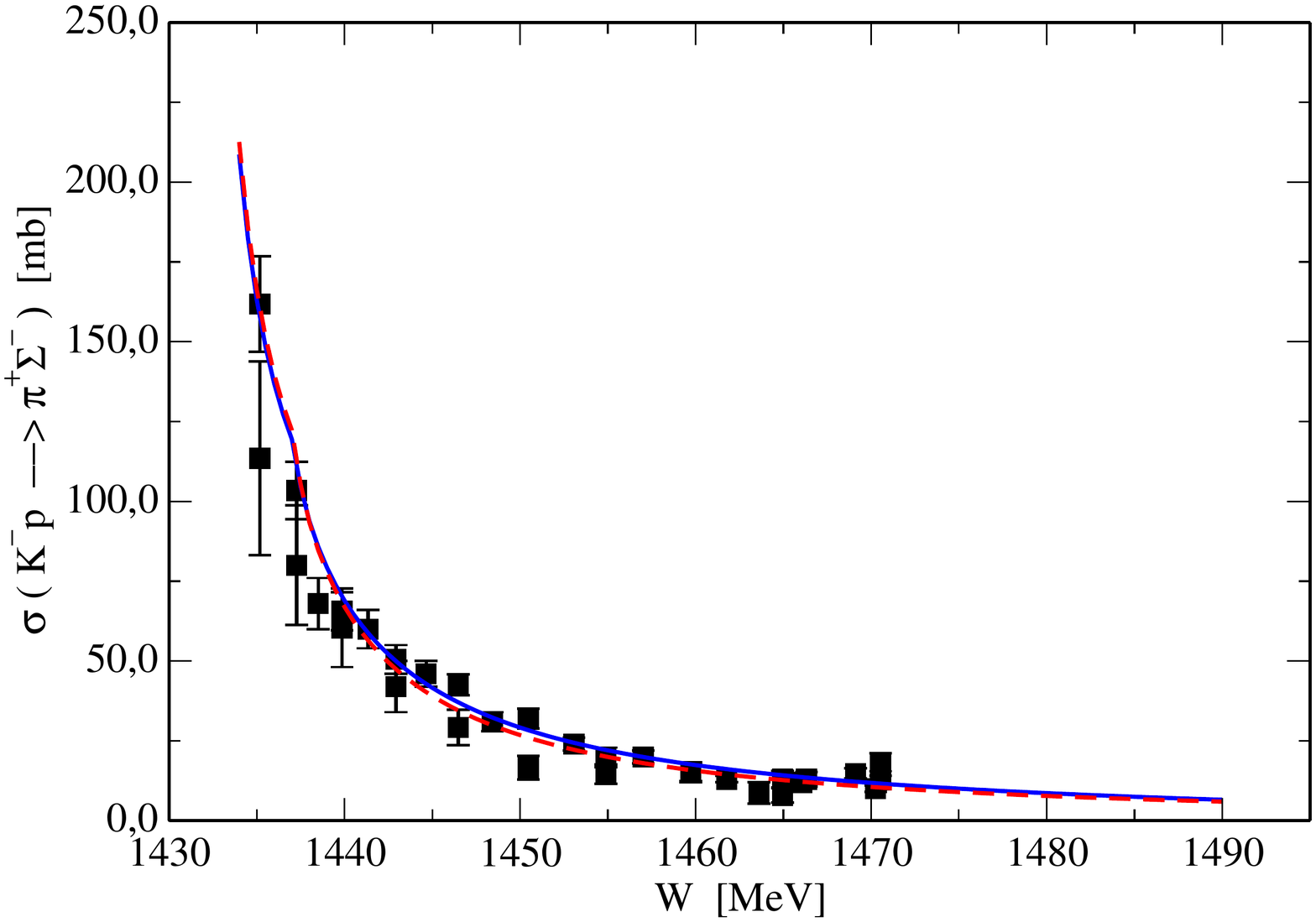} \\[-3mm]
\includegraphics[width=0.42\textwidth]{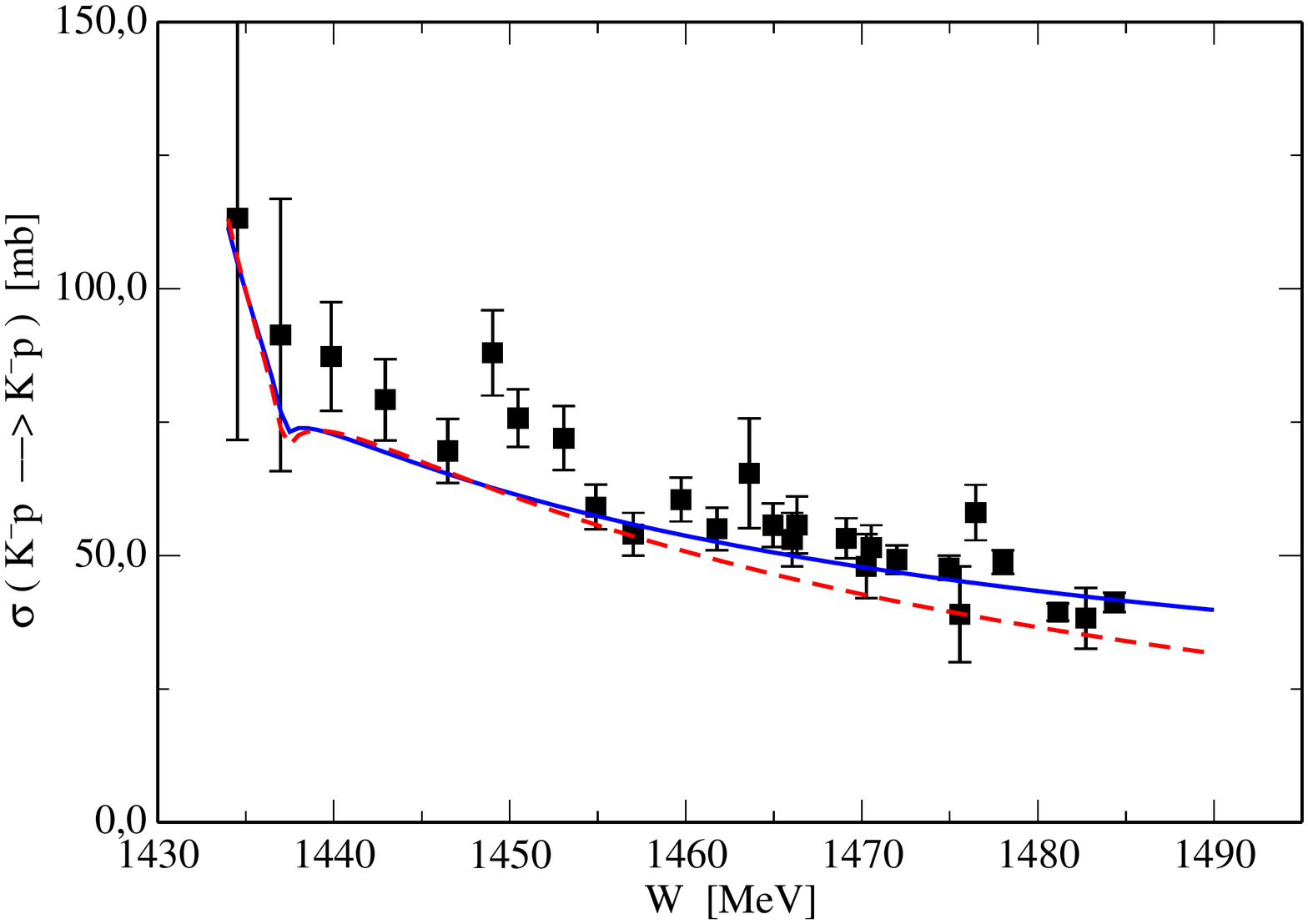}
\includegraphics[width=0.42\textwidth]{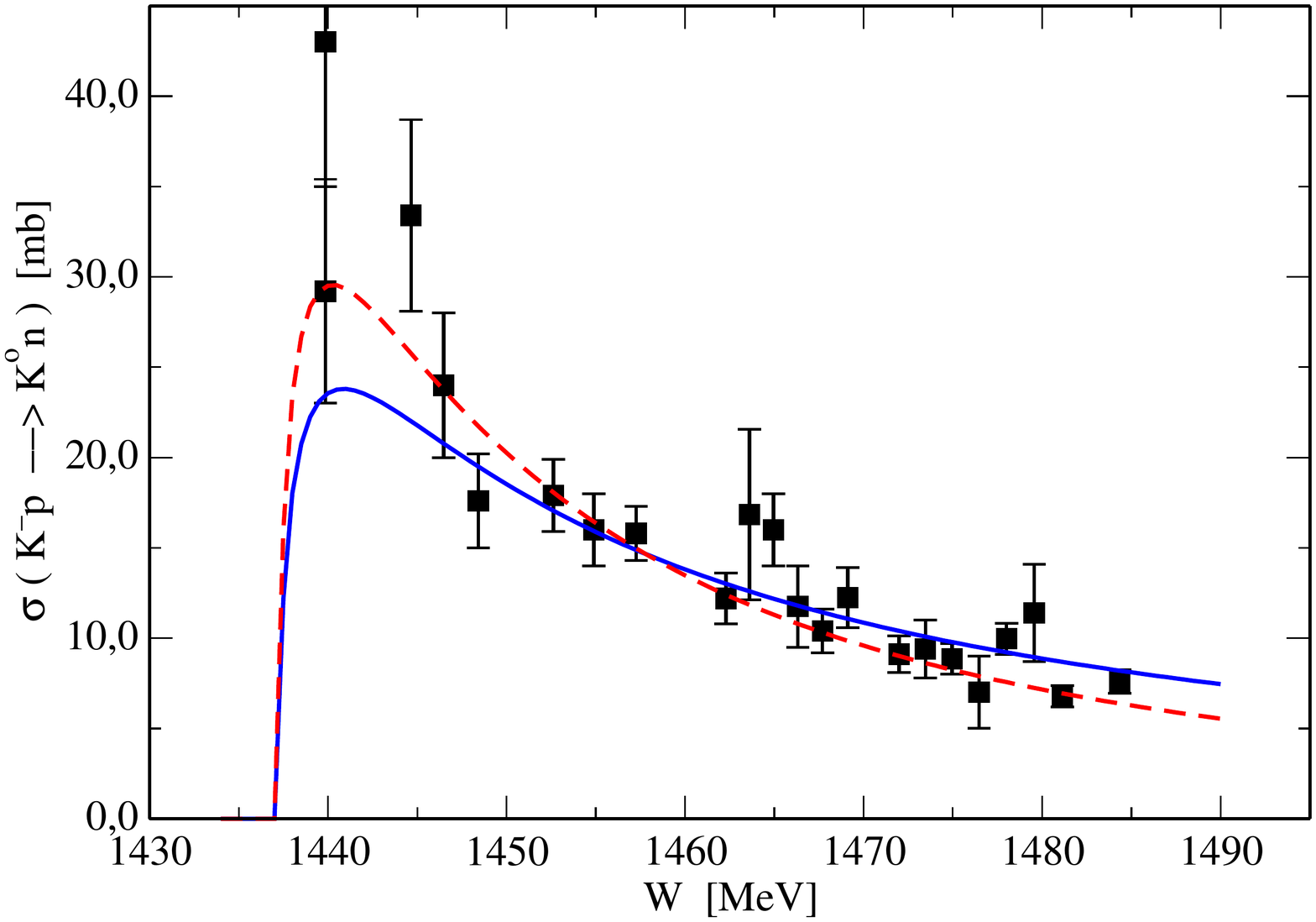} \\[-3mm]
\includegraphics[width=0.42\textwidth]{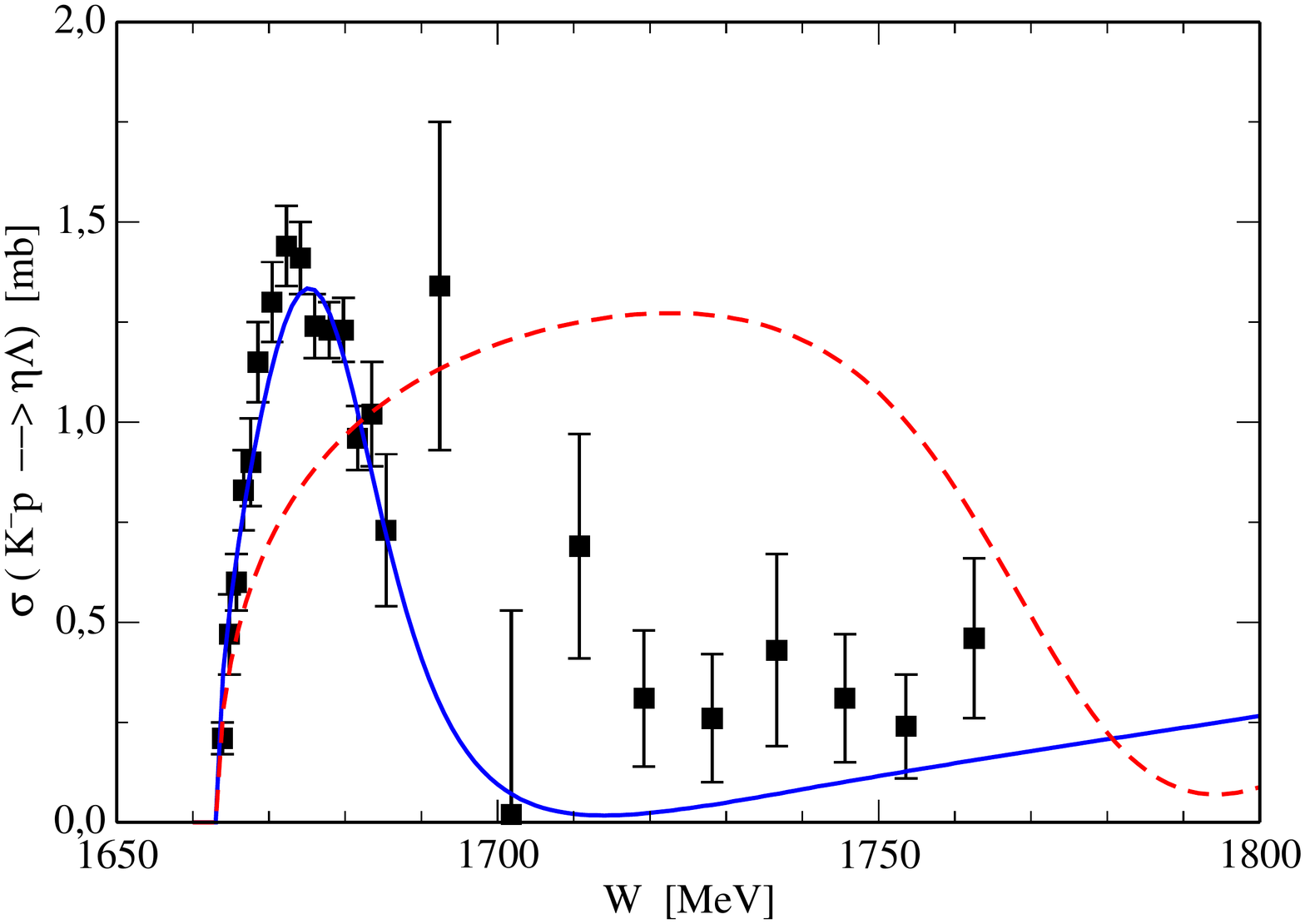}
\includegraphics[width=0.42\textwidth]{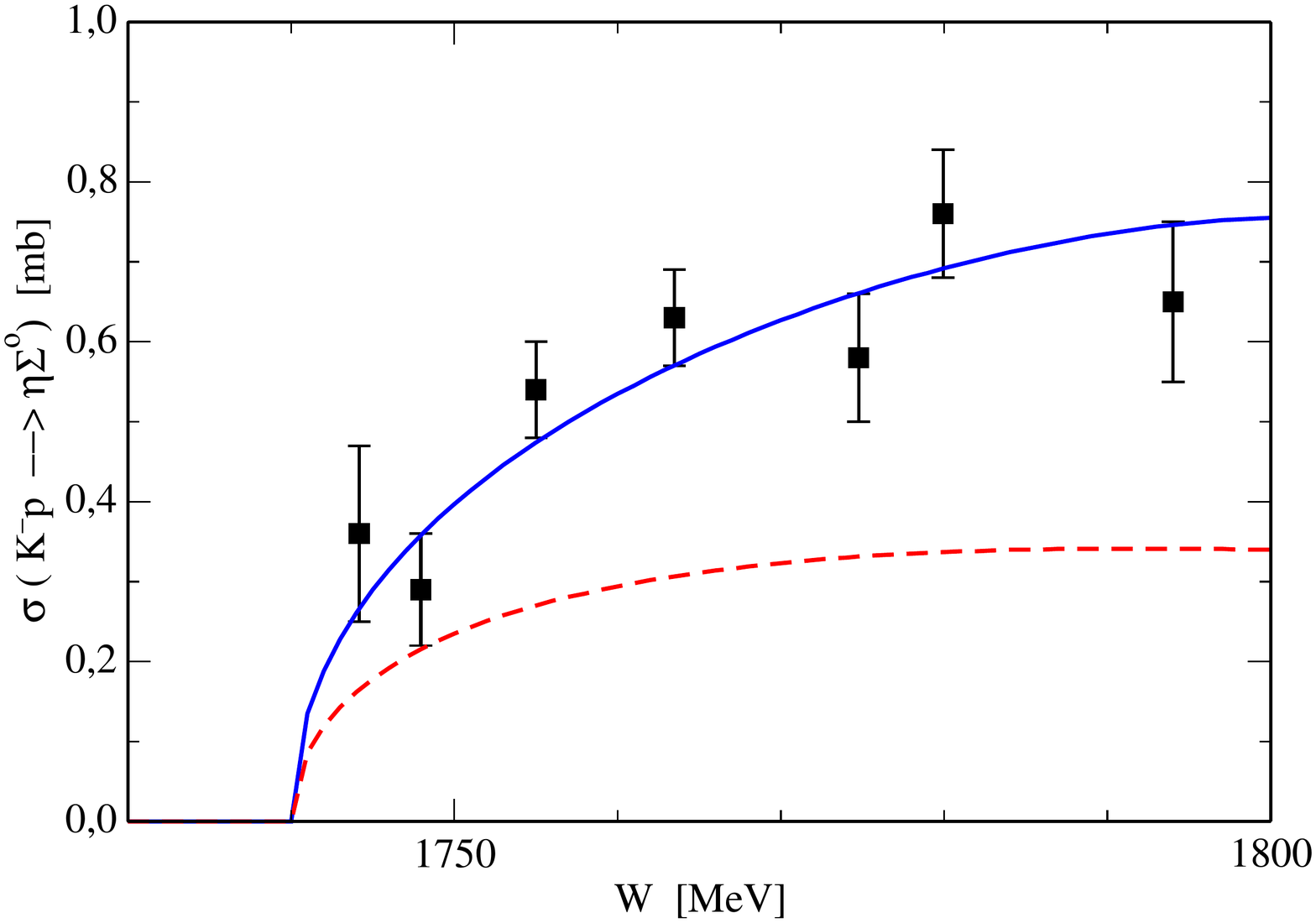}
\caption{The $K^{-}p \rightarrow MB$ total cross sections. Continuous blue lines show our current 
model predictions, the dashed red lines the predictions of the NLO30 Prague model.
}
\label{fig:xsec}
\end{figure}

Our reproduction of the fitted cross sections (as a function of the MB c.m.~energy $W=\sqrt{s}$) is shown in Figure~\ref{fig:xsec} in comparison 
with the predictions of the Prague NLO30 model \cite{Cieply:2011nq}. As the experimental data 
come from relatively old bubble chamber measurements, they are not very precise and the $\bar{K}N$ 
models have no problem to reproduce them. Though, we note that the NLO30 model is missing 
completely with regards to the $\eta \Lambda$ and $\eta \Sigma$ cross sections that were not included 
in the respective fit. There, our current model does really well when reproducing the $\Lambda(1670)$ 
structure observed above the channel threshold in the $\eta \Lambda$ cross section. 

%..............................................................................................
%..............................................................................................

The branching ratios measured at the $\bar{K}N$ threshold and the kaonic hydrogen characteristics 
put much tighter restrictions on the theoretical models. In Table~\ref{tab:fits} we show how 
these data are reproduced by the current and the NLO30 models. The same table also presents 
the model predictions for the poles of the scattering amplitude generated in the $I=0$ sector. 
The $z_1$ and $z_2$ poles are found on the $[-,+,+,+]$ Riemann sheet (RS) and are both assigned 
to the $\Lambda(1405)$ resonance. The $z_3$ pole appears on the $[-,-,-,+]$ RS and is related 
to the $\Lambda(1670)$ resonance. As we already mentioned the NLO30 model did not include 
the cross sections data available at the higher energies, so it is not surprising that its prediction 
of the $\Lambda(1670)$ pole position is completely off the mark. However, it is worth noting that even 
in this case a third isoscalar pole is still predicted by the chirally motivated coupled channels 
model, and inclusion of relevant experimental data brings the pole position in a good agreement 
with the PDG accounts on the $\Lambda(1670)$. Thus, we conclude that both discussed models 
predict three dynamically generated $\Lambda^*$ states, a feature which is completely in line 
with results provided by other $\bar{K}N$ approaches based on chiral Lagrangians. 
The two-pole structure of $\Lambda(1405)$ is a typical feature of coupled-channel models 
based on chiral symmetry \cite{Hyodo:2011ur,Bruns:2019bwg,Meissner:2020khl,Mai:2020ltx}.

%----------------------------------------------------------------------------------------------
\begin{table}[htb!]
\caption{Fitted branching ratios and characteristics of kaonic hydrogen in comparison
    with the experimental data. The positions and assignment of dynamically generated 
    $\Lambda^{*}$ resonances are shown as well.}
\begin{center}
\begin{tabular}{c|cc|c}
\toprule \toprule
                  & current fit & NLO30 fit & experiment \\ \midrule \midrule
 $\gamma$         & 2.36  & 2.37  & 2.36(4)   \\
 $R_c$            & 0.656 & 0.660 & 0.664(11) \\
 $R_n$            & 0.207 & 0.191 & 0.189(15) \\ \midrule
 $\Delta E_N(1s)$ [eV] &  307  &  310  &  283(42)  \\
 $\Gamma (1s)$ [eV]    &  590  &  607  &  541(111) \\ \midrule
 $z_1$ [MeV] & (1353,-43) & (1355,-86)  & $\Lambda_1(1405)$ \\
 $z_2$ [MeV] & (1429,-24) & (1418,-44)  & $\Lambda_2(1405)$ \\
 $z_3$ [MeV] & (1677,-14) & (1774,-35)  & $\Lambda(1670)$ \\
\bottomrule
\end{tabular}
\end{center}
\label{tab:fits}
\end{table}
%----------------------------------------------------------------------------------------------

\begin{figure}[htb!]    
\centering
\includegraphics[width=0.48\textwidth]{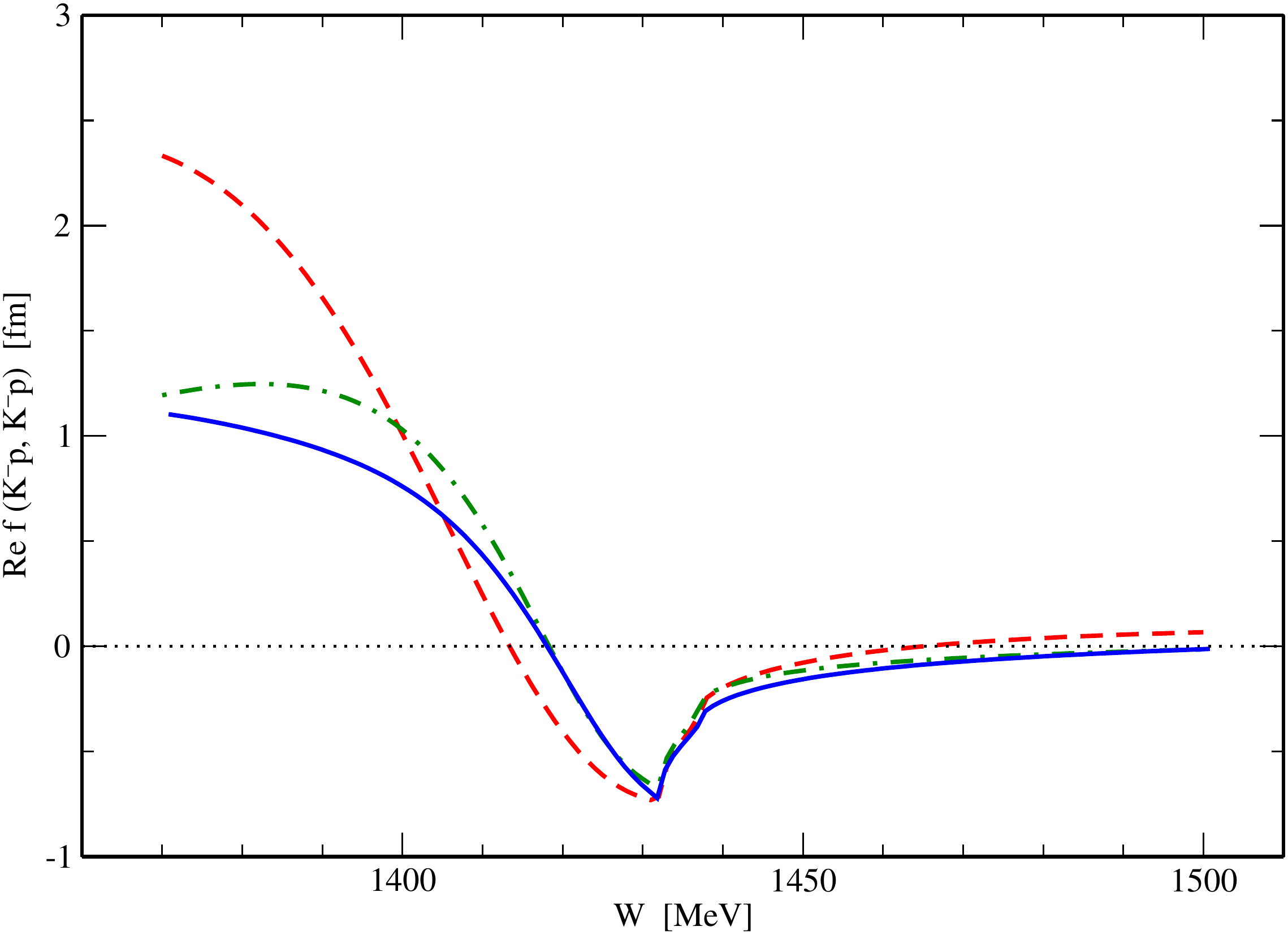} \hspace*{2mm}
\includegraphics[width=0.48\textwidth]{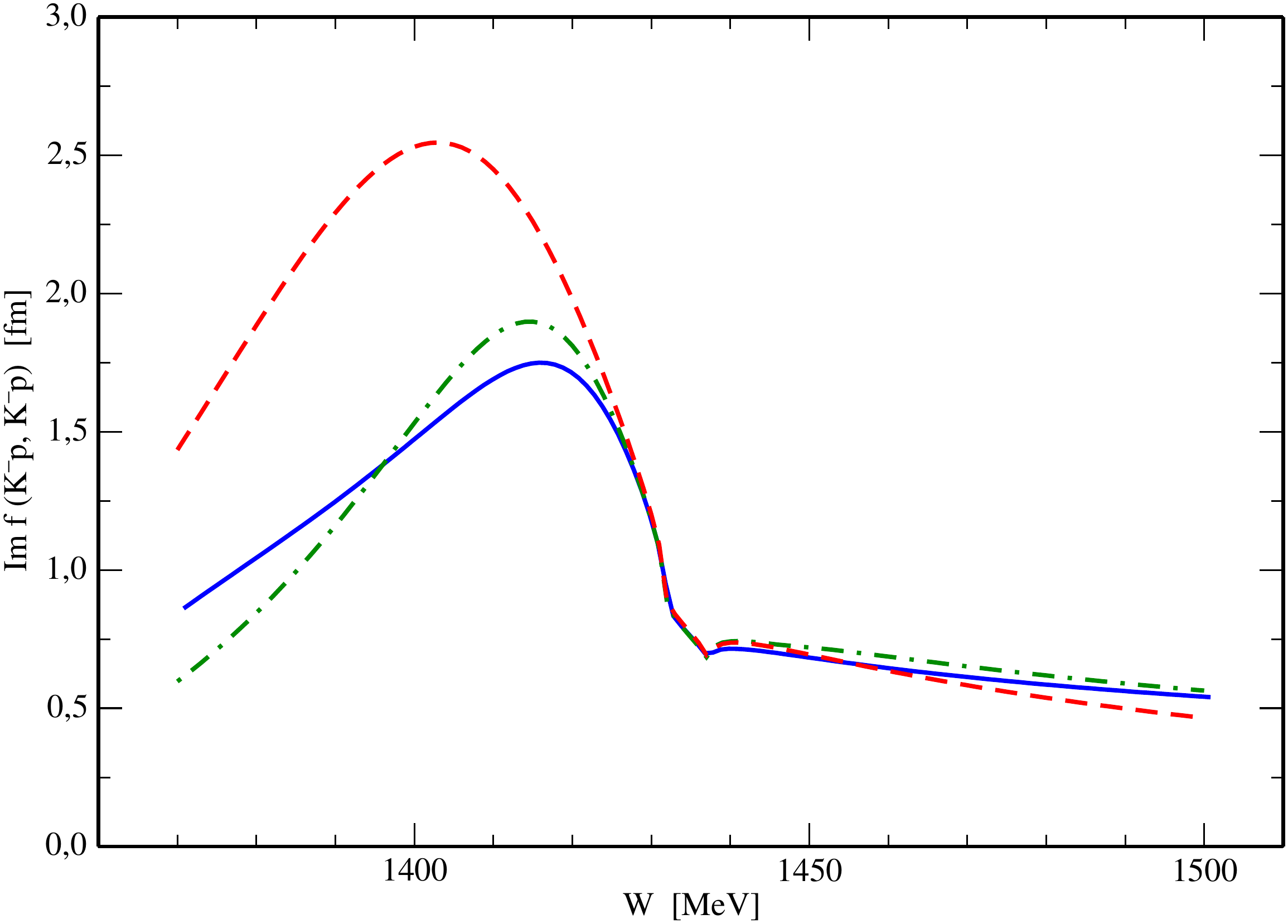} 
\caption{Energy dependence of the $K^{-}p$ amplitude. The real and imaginary parts of the amplitude 
are shown in the left and right panels, respectively. Continuous blue lines represent our current 
model predictions, the dashed red lines the predictions of the NLO30 Prague model, 
and the dot-dashed green lines the predictions of the Kyoto-Munich model.
}
\label{fig:Kp}
\end{figure}

We conclude the presentation of our new $\bar{K}N$ model by demonstrating the energy dependence 
of the $K^{-}p$ elastic amplitude. This is done in Fig.~\ref{fig:Kp} where our results are shown 
in comparison with those generated by the NLO30 model and the Kyoto-Munich model \cite{Ikeda:2012au}. 
Interestingly, our current model provides the $K^{-}p$ energy dependence very similar to the one 
obtained by the latter approach. Considering the fact that the higher energy cross sections were not 
included in the analysis performed in \cite{Ikeda:2012au} and there are more instances where 
the two models differ (e.g.~by a use of separable potentials in our approach) the agreement 
of the two models seems to be rather coincidental.

%..............................................................................................
%..............................................................................................

%\newpage
%%%%%%%%%%%%%%%%%%%%%%%%%%%%%%%%%%%%%%%%%%%%%%%%%%%%%%%%%%%%%%%%%%%%%%%%%%%%%%%%%%%%%%%%%%%%%%%
\section{SU(3) flavor symmetry in coupled-channel models}
\label{sec:SU3}
%%%%%%%%%%%%%%%%%%%%%%%%%%%%%%%%%%%%%%%%%%%%%%%%%%%%%%%%%%%%%%%%%%%%%%%%%%%%%%%%%%%%%%%%%%%%%%%

In what follows we consider only the fully flavor-symmetric case of s-wave $S=-1$ MB 
scattering, where the meson and the baryon are members of the respective ground-state octets. 
From the tensor product decomposition of multiplets, 
$\bm{8}\otimes\bm{8}=\bm{1}\oplus\bm{8}\oplus\bm{8}\oplus\bm{10}\oplus\bm{\overline{10}}\oplus\bm{27}$, 
one expects $s$-channel exchanges of singlet, octet, decuplet and $27$-plet resonances in such a process. 
As there are no decuplet states with charge $Q=0$, $S=-1,\,I=L=0$, we can disregard them for our purposes. 
The relevant MB states for our study are (see Appendix~B of \cite{Bruns:2019fwi} for our phase conventions)
\begin{eqnarray}
|\pi\Sigma\rangle_{I=0} &=& -\frac{1}{\sqrt{3}}\left(|\pi^{0}\Sigma^{0}\rangle + |\pi^{-}\Sigma^{+}\rangle + |\pi^{+}\Sigma^{-}\rangle\right)\,,\nonumber \\
|\bar{K}N\rangle_{I=0} &=& -\frac{1}{\sqrt{2}}\left(|\bar{K}^{0}n\rangle + |K^{-}p\rangle\right)\,,\quad |\eta\Lambda\rangle_{I=0} = |\eta\Lambda\rangle\,,\nonumber \\
|K\Xi\rangle_{I=0} &=& \frac{1}{\sqrt{2}}\left(|K^{0}\Xi^{0}\rangle + |K^{+}\Xi^{-}\rangle\right)\,.\label{eq:iso0}
\end{eqnarray}
From these one can construct the following set of MB flavor eigenstates:
\begin{eqnarray}
  |(\bm{1})\rangle_{I=0} %&=& \frac{1}{\sqrt{8}}\left(|\pi^{0}\Sigma^{0}\rangle + |\pi^{-}\Sigma^{+}\rangle + |\pi^{+}\Sigma^{-}\rangle + |\bar{K}^{0}n\rangle + |K^{-}p\rangle + |\eta\Lambda\rangle + |K^{0}\Xi^{0}\rangle + |K^{+}\Xi^{-}\rangle\right) \nonumber \\
  \!&=&\! \frac{1}{\sqrt{8}}\left(|\eta\Lambda\rangle_{I=0} -\sqrt{2}|\bar{K}N\rangle_{I=0} + \sqrt{2}|K\Xi\rangle_{I=0} -\sqrt{3}|\pi\Sigma\rangle_{I=0}\right),\nonumber\\
|(\bm{8})\rangle_{I=0}  \!&=&\! \frac{1}{\sqrt{10}}\left(\sqrt{2}|\eta\Lambda\rangle_{I=0} -|\bar{K}N\rangle_{I=0} + |K\Xi\rangle_{I=0} +\sqrt{6}|\pi\Sigma\rangle_{I=0}\right),\label{eq:O1state}\\
|(\bm{8'})\rangle_{I=0} \!&=&\! \frac{1}{\sqrt{2}}\left(|\bar{K}N\rangle_{I=0} + |K\Xi\rangle_{I=0}\right),\label{eq:O2state}\\
|(\bm{27})\rangle_{I=0} \!&=&\! \frac{1}{\sqrt{40}}\left(3\sqrt{3}|\eta\Lambda\rangle_{I=0} +\sqrt{6}|\bar{K}N\rangle_{I=0} - \sqrt{6}|K\Xi\rangle_{I=0} -|\pi\Sigma\rangle_{I=0}\right).\nonumber %\label{eq:27state}
\end{eqnarray}
A quick and straightforward way to obtain symmetry constraints for the couplings of possible $\Lambda^{\ast}$ resonances 
to the above states is given by writing down effective chiral Lagrangians containing {\it explicit\,} fields pertaining 
to the $\Lambda^{\ast}$ states. For a singlet and an octet $\Lambda^{\ast}$, respectively, one can adopt
\begin{eqnarray}
  \mathcal{L}_{Y}^{(\bm{1})} &=& \frac{D_{S}}{2}\langle\bar{B}\gamma^{\mu}u_{\mu}\rangle \,\Lambda^{\ast}\,+\,h.c.\,,\\
  \mathcal{L}_{Y}^{(\bm{8})} &=& \frac{D_{Y}}{2}\langle\bar{B}\gamma^{\mu}\lbrace u_{\mu},\,Y\rbrace\rangle + \frac{F_{Y}}{2}\langle\bar{B}\gamma^{\mu}\lbrack u_{\mu},\,Y\rbrack\rangle\,+\,h.c.\,,
\end{eqnarray}
where $Y=\frac{1}{\sqrt{6}}\mathrm{diag}\left(1,\,1,\,-2\right)\Lambda^{\ast}+(\mathrm{other\,\,fields})\,$, $u^{\mu}$ and $B$ contain the meson and baryon octet fields in the usual fashion (compare e.g.~Appendix~A of \cite{Bruns:2019fwi} for our notation), and $D_{S}$, $D_{Y}$, $F_{Y}$ are coupling constants.  We shall not encounter a $27$-plet resonance, so we omit a corresponding effective Lagrangian here (see, however, Eq.~(4.24) in \cite{Oh:2004gz}). For a later reference, we also provide an example of an application of the above Lagrangians: From $\mathcal{L}_{Y}^{(\bm{1})}$, one derives the tree-level contribution to the $I=0$ s-wave amplitude for $\pi\Sigma$ scattering due to an $s$-channel exchange of a singlet $\Lambda^{\ast}$ hyperon as 
\begin{equation}\label{eq:treegraphLstar}
    f_{0+,I=0}^{(\mathrm{tree})}(\pi\Sigma\rightarrow\pi\Sigma) = -\frac{3D_{S}^2}{2F_{\pi}^2}\frac{(E_{\Sigma}+m_{\Sigma})}{8\pi\sqrt{s}}\frac{(\sqrt{s}-m_{\Sigma})^2}{\sqrt{s}-m_{\Lambda^{\ast}}}\,.
\end{equation}
We do not incorporate such graphs with explicit resonances in our model, but only use them 
to illustrate some general features of the resonance-pole terms from a different viewpoint. 
Specifically, we focus on ratios of moduli of couplings $\beta$ 
that determine how strongly the MB channels couple to a specific $\Lambda^*$ state, see Eq.~(\ref{eq:res}) below as well. 
We anticipate that in these ratios the conventional prefactors and phases drop out. With the help 
of the above formalism, for the singlet case one deduces
\begin{eqnarray}
r_{12} &:=& \left|\frac{\beta(\Lambda^{\ast}\rightarrow\pi\Sigma)}{\beta(\Lambda^{\ast}\rightarrow \bar{K}N)}\right| = \sqrt{\frac{3}{2}}\,,\quad r_{13} := \left|\frac{\beta(\Lambda^{\ast}\rightarrow\pi\Sigma)}{\beta(\Lambda^{\ast}\rightarrow \eta\Lambda)}\right| = \sqrt{3}\,, \nonumber \\ r_{24} &:=& \left|\frac{\beta(\Lambda^{\ast}\rightarrow\bar{K}N)}{\beta(\Lambda^{\ast}\rightarrow K\Xi)}\right| = 1\,,
\label{eq:r1}
\end{eqnarray}
while for an octet $\Lambda^{\ast}$, the corresponding ratios are of the form
\beq
r_{12}  = \left|\frac{D_{Y}}{D_{Y}+3F_{Y}}\right|\,,\quad r_{13}  = \sqrt{3}\,,\quad r_{24}  = \left|\frac{D_{Y}+3F_{Y}}{D_{Y}-3F_{Y}}\right|\,,
\eeq{eq:r8}
and finally for a 27-plet member,
\beq
r_{12}  = \frac{1}{\sqrt{6}}\,,\quad r_{13}  = \frac{1}{3\sqrt{3}}\,,\quad r_{24}  = 1\,.
\eeq{eq:r27}
Relations like this are, of course, well-known - we compile them here just for an easy comparison 
with our results presented in the next section. In \ref{app:traj}, we collect the formulae used 
 to map out the trajectory from the {\it real world} setting to a conveniently chosen symmetric one. 
Before we go on to apply the above analysis to our coupled-channel formalism, we have to make a remark 
on the octet states. In general, a $\Lambda^{\ast}$ octet state will couple to a mixture of the two octet 
states introduced in Eqs.~(\ref{eq:O1state}-\ref{eq:O2state}), parameterized by an angle $\theta$. 
Taking this into account, we expect one of
\begin{eqnarray*}
  |(\bm{1})\rangle &\equiv& |(\bm{1})\rangle_{I=0}\,,\quad |(\bm{27})\rangle \equiv |(\bm{27})\rangle_{I=0}\,,\\
  |(\bm{8a})\rangle &=& \cos\theta\,|(\bm{8})\rangle_{I=0} - \sin\theta\,|(\bm{8'})\rangle_{I=0}\,,\\
  |(\bm{8b})\rangle &=& \sin\theta\,|(\bm{8})\rangle_{I=0} + \cos\theta\,|(\bm{8'})\rangle_{I=0}\,,
\end{eqnarray*}
to be the MB state coupled to an $s$-channel resonance generated in our framework. More generally, 
in case of exact flavor symmetry the $4\times 4$ coupled channel matrix $v$ containing 
our interaction kernel must be diagonalized by an orthogonal matrix of the form
\begin{equation}\label{eq:Strafo}
  \mathcal{S} = \left(\begin{array}{c|c|c|c} -\sqrt{\frac{3}{8}}\,\, & \sqrt{\frac{6}{10}}\cos\theta & \sqrt{\frac{6}{10}}\sin\theta & \,-\frac{1}{\sqrt{40}} \\
    -\sqrt{\frac{2}{8}}\,\, & -\frac{\cos\theta}{\sqrt{10}} - \frac{\sin\theta}{\sqrt{2}} & -\frac{\sin\theta}{\sqrt{10}} + \frac{\cos\theta}{\sqrt{2}}& \,\sqrt{\frac{6}{40}} \\
      \frac{1}{\sqrt{8}}\,\, & \sqrt{\frac{2}{10}}\cos\theta & \sqrt{\frac{2}{10}}\sin\theta & \,3\sqrt{\frac{3}{40}} \\
      \sqrt{\frac{2}{8}}\,\, & \frac{\cos\theta}{\sqrt{10}} - \frac{\sin\theta}{\sqrt{2}} & \frac{\sin\theta}{\sqrt{10}} + \frac{\cos\theta}{\sqrt{2}}& \,-\sqrt{\frac{6}{40}}\end{array}\right)\,,\quad %\mathcal{S}^{\top}\mathcal{S}=\mathcal{S}\mathcal{S}^{\top}=\mathds{1}_{4\times4}\,,
\end{equation}
i.e.~$v=\mathcal{S}v_{\mathrm{diag}}\mathcal{S}^{\top}$, where
$v_{\mathrm{diag}}=\mathrm{diag}(v_{(\bm{1})},\,v_{(\bm{8a})},\,v_{(\bm{8b})},\,v_{(\bm{27})})$. 
In the generic case, when $v_{(\bm{8a})} \not = v_{(\bm{8b})}$, it should be possible to determine 
the angle $\theta$ (at a given energy) from this requirement. However, for $v_{(\bm{8a})} = v_{(\bm{8b})}$, 
e.g.~for a pure WT kernel, this is not possible. In this particular case the degeneracy appears due 
to an {\it accidental symmetry} of the WT kernel. 
Since our full scattering amplitude is given in a form of a geometric series, $t=v+vGv+vGvGv+\ldots$,
composed from the interaction kernel $v$ and the diagonal matrix $G$ containing only one universal 
loop function for all four channels in the SU(3)-symmetric case ($G\sim\mathds{1}_{4\times4}$), it is clear that 
$t=\mathcal{S}(\mathrm{diag}(t_{(\bm{1})},\,t_{(\bm{8a})},\,t_{(\bm{8b})},\,t_{(\bm{27})}))\mathcal{S}^{\top}$ too. 
Thus, we can extract the coupling ratios 
for a given resonance (or a bound state) from the pertinent pole residues of the $t$-matrix, 
and classify the resonances uniquely according to their flavor content.

In the real world, the flavor symmetry is broken and, unfortunately, things are  not so simple. 
Even if the flavor-breaking effects in the interaction kernel are small, the loop functions will be 
different in each channel, and we have to deal with $2^{n}$ Riemann sheets (RSs) of the complex energy 
surface (for $n$ channels), where the $G$ entries can be different depending on their threshold energies. 
This will introduce a {\it kinematical flavor breaking} for resonances located on a sheet with mixed signature, 
i.e.~with different signs of the imaginary parts of the various loop functions. Thus, the symmetry constraints 
discussed above will not apply to a pole located on a RS with a mixed signature, even in the limit 
of zero flavor breaking. However, as the interval on the physical real axis connecting the {\it mixed} 
sheet and the physical sheet shrinks to a point when this limit is approached, the resonance also becomes 
{\it unphysical} (i.e. unobservable). In this way, an apparent contradiction between measurable resonance 
properties and exact flavor symmetry is avoided. We shall see examples of such behavior in the next section. 
In addition, even if a resonance pole ends up on the physical RS (as a bound state), or the unphysical RS 
connected to it in the symmetry limit, it can switch over from one sheet to another for a certain value 
of the flavor breaking, in which case the coupling ratios can vary quite dramatically. While it is hard 
to judge how {\it realistic} such scenarios are, these observations point to the fact that a smooth dependence 
on the flavor breaking, and a simple connection to the fully SU(3)-symmetric scenario, is not guaranteed 
for all resonances generated in a multi-channel system.

%%%%%%%%%%%%%%%%%%%%%%%%%%%%%%%%%%%%%%%%%%%%%%%%%%%%%%%%%%%%%%%%%%%%%%%%%%%%%%%%%%%%%%%%%%%%%%%
\section{Results and discussion}
\label{sec:results}
%%%%%%%%%%%%%%%%%%%%%%%%%%%%%%%%%%%%%%%%%%%%%%%%%%%%%%%%%%%%%%%%%%%%%%%%%%%%%%%%%%%%%%%%%%%%%%%

In this section we discuss how the broken SU(3) flavor symmetry affects the properties 
of dynamically generated resonances that are observed in the isoscalar $S=-1$ sector.
The chirally motivated coupled channel model introduced in Sec.~\ref{sec:model} generates two 
resonant states assigned to $\Lambda(1405)$ and another one to the $\Lambda(1670)$ resonance, 
see Table \ref{tab:fits} for the positions of the pertinent poles at complex energies $z = \sqrt{s}$. 
These poles are located at RSs that would be called {\it the second RS} if only two channels 
were considered, e.g.~on the RS that is connected with the physical region at the real axis around 
the given pole energy ${\rm Re}\:z$. In our notation, they are the $[-,+,+,+]$ and $[-,-,-,+]$ RS 
for the $\Lambda(1405)$ and the $\Lambda(1670)$ poles, respectively. We remark that the appearance 
of the resonances on the RSs that are not completely symmetric with respect to all coupled channels 
is a consequence of the broken SU(3) flavor symmetry in the physical limit. As we will see 
one can gradually restore the SU(3) symmetry and track the movement of the poles. However, 
as we already indicated at the end of the previous section, it is not guaranteed that the 
pole will reside on a channel symmetric RS (either all channel momenta physical or all 
of them unphysical) when one reaches the SU(3) flavor limit this way.

The movement of the poles from (or to) the SU(3) flavor limit was already explored in \cite{Jido:2003cb}. 
In the present work we follow a similar procedure by introducing a scaling parameter $x$ 
but opt for a slightly different approach to the variations of the hadron masses and other model parameters 
(decay constants $F_j$ and inverse ranges $\alpha_{jb}$) when defining the trajectory from 
the limit of restored SU(3) flavor symmetry (at $x = 0$) to the physical limit (at $x = 1$), 
see \ref{app:traj} for details. When going from the physical limit by restoring gradually 
the SU(3) flavor symmetry we follow the movement of the poles as well as the ratios 
of the channel couplings $r_{ij}$ introduced in Eqs.~(\ref{eq:r1}-\ref{eq:r27}) in the previous section.
The couplings $\beta_{jb}$ that are required for the ratios can be calculated as residua 
of the scattering amplitude, 
\beq
  {\rm Res}_{z=z_R}f_{ia,jb}(z) = \beta_{ia}\,\beta_{jb}\; .
\eeq{eq:res}
Referring back to the example given in Eq.~(\ref{eq:treegraphLstar}), we do not expect any phase-space factors $\sim q(s)$ contained in the couplings $\beta$. 
The coupling factors are also related to the {\it compositeness} $\cal{C}$ of the dynamically generated state assigned 
to the pertinent pole \cite{Baru:2003qq,Gamermann:2009uq,Aceti:2012dd,Sekihara:2014kya} via
\beq
  {\cal C}_{jb}(z_R) = \beta_{jb}^{\:2} \:\frac{dG_{jb}}{dz}(z_R)\; ,
\eeq{eq:composit}
where the loop function $G(z)$ is the one introduced in Eq.~(\ref{eq:f0pMOD}).

%..............................................................................................
\begin{figure}[htb!]    
\centering
\includegraphics[width=0.75\textwidth]{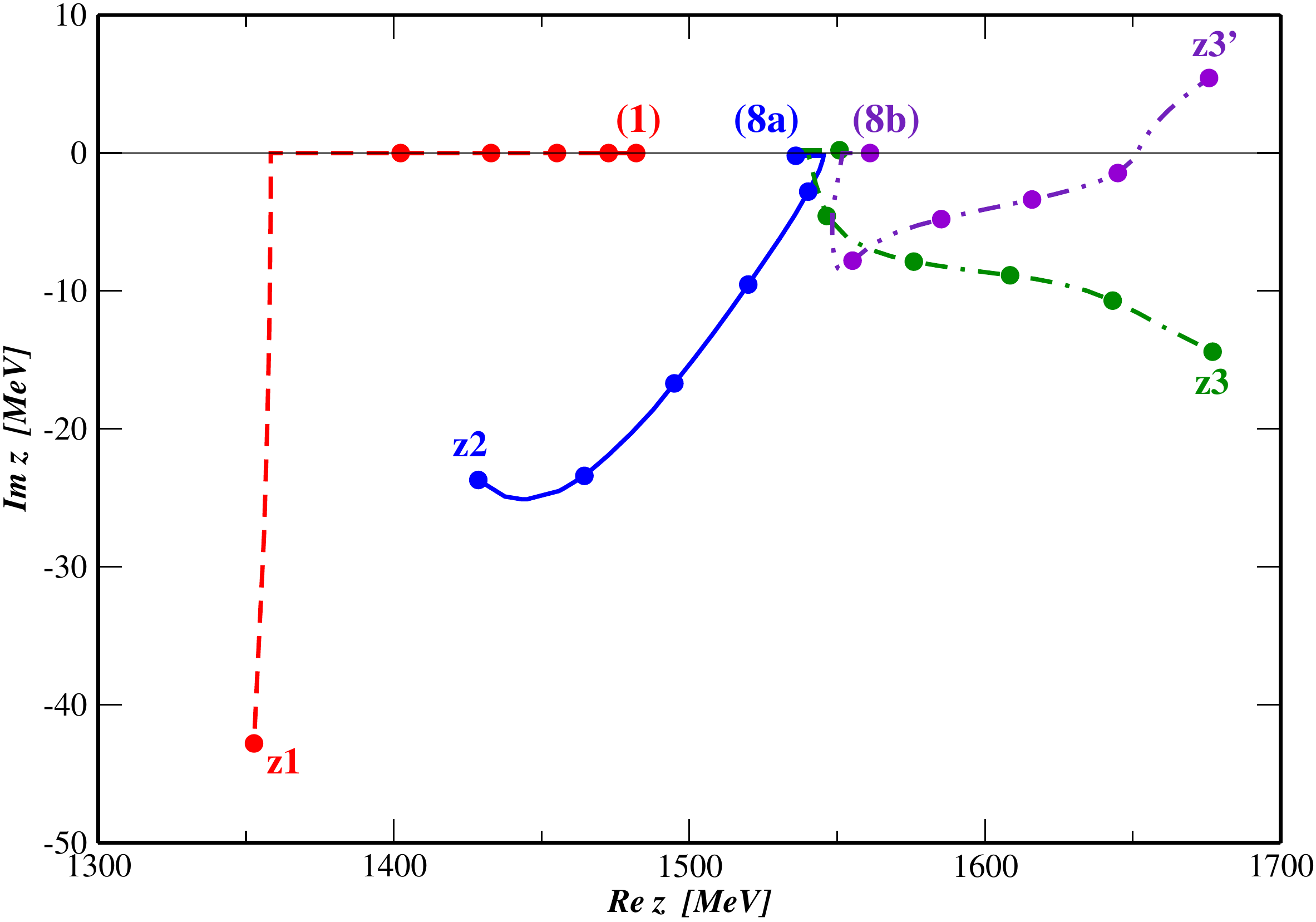} 
\caption{Movements of the poles upon varying the $x$ parameter that scales the breaking 
of the SU(3) flavor symmetry. The pole positions in the physical limit ($x=1$) are marked 
as $z1$, $z2$ and $z3$, the circles on the pole trajectories indicate the positions of the poles 
for $x$ values modified in steps of $0.2$. See text for explanation on the shadow $z3'$ 
pole and more details.
}
\label{fig:poles}
\end{figure}
%..............................................................................................

At first, let us have a look at the movement of the poles as one goes from the physical limit 
to the point of the restored SU(3) flavor symmetry. The pole trajectories are shown in Fig.~\ref{fig:poles} 
and the starting positions (for $x=1$) of the $z_1$, $z_2$ and $z_3$ poles are exactly those 
given in Table \ref{tab:fits}. As we gradually decrease the scaling parameter $x$ from 1 to 0 
the poles move along their respective paths reaching (for $x=0$) the end points located 
at the real axis. We also mark by full circles the pole positions at $x$ values that 
are multiples of 0.2, so one can judge how fast the poles move. The trajectory of the $z_1$ 
pole (the {\it lower mass} $\Lambda_1(1405)$ pole) is visualized by the red dashed line 
in the figure. As the $x$ value is reduced the pole moves very fast to reach the real 
axis for $x = 0.9$, then continues to move along it to reach its final position 
at the energy $z_{\bm{(1)}} = 1482$ MeV, a point we identify as a SU(3) flavor singlet. As the $z_1$ 
pole moves along the real axis it also switches from the $[-,+,+,+]$ RS to the $[+,+,+,+]$ RS 
which happens for $x = 0.85$ when the pole catches on the lowest $\pi \Sigma$ threshold 
(that also keeps moving when we vary $x$). The movement of the $z_2$ pole 
(the {\it higher mass} $\Lambda_2(1405)$ one) is visualized by the continuous blue line and 
we find it more fluid with a major part of its trajectory on the $[-,+,+,+]$ RS. 
The pole reaches the real axis for $x = 0.11$, immediately hits the lowest threshold 
and switches to the physical $[+,+,+,+]$ RS ending its movement (for $x=0$) 
at the energy $z_{\bm{(8a)}} = 1536$ MeV, a point we associate with one of the SU(3) octets. Thus, each of 
the two $\Lambda(1405)$ poles goes to a different state in the SU(3) flavor limit, both states 
located on the physical RS (in all channels).

Although the $\Lambda(1670)$ related $z_3$ pole trajectory in Fig.~\ref{fig:poles}, 
(see the green dot-dashed line there) is quite similar to the $z_2$ pole movement 
the situation is more complicated in this case. The pole spends most of its movement 
on the $[-,-,-,+]$ RS, reaches the real axis at $x = 0.15$, slightly below the 
lowest $\pi\Sigma$ threshold but when it hits this threshold it ends up on the $[+,-,-,+]$ RS. 
Although the pole movement ends on the real axis, the RS is not channel symmetric, 
so the final point of the $z_3$ pole trajectory cannot be identified with any SU(3) flavor 
symmetric state. In fact, we were able to locate the second SU(3) octet pole at the (all 
channels physical) $[+,+,+,+]$ RS at the energy  $z_{\bm{(8b)}} = 1561$ MeV. Thus, the $z_3$ pole 
movement apparently aims at this {\bf (8b)} state, but ends at the energy $z_3(x=0) = 1551$ MeV, 
10 MeV below the {\bf (8b)} state and on a different RS. 

Of course, the pole movement can also be followed in the opposite way, starting from the pole position 
found for $x=0$ and increasing gradually the $x$ value. When one does so for the {\bf (8b)} 
octet state we get the pole trajectory shown by the magenta dot-dot-dashed line in Fig.~\ref{fig:poles}. 
The pole movement starts on the physical $[+,+,+,+]$ RS, then switches to the $[-,+,+,+]$ RS, leaves 
the real axis, and finally crosses the real axis in between the (moving) $\eta\Lambda$ and $K\Xi$ 
thresholds switching to the $[+,-,-,+]$ RS, where it ends its movement at the complex energy 
$z'_3 = (1676, 5.4)$ MeV. We have checked the origin of this pole by looking at its movement 
to the zero coupling limit (ZCL) in which the inter-channel couplings are switched off. In this limit,
both the $z_3$ and $z'_3$ poles coalesce and are represented by the same $K\Xi$ bound state, 
thus the two poles originate from the same ZCL state and represent shadow poles to each other 
in the physical limit, see Refs.~\cite{Eden:1964zz}, \cite{Pearce:1988rk} and \cite{Cieply:2016jby} 
for more details on the the ZCL and the concept of shadow poles. In the physical limit, 
the $z_3$ and $z'_3$ poles are both located at very similar energies (do note they both have 
partners at complex conjugated energies) but since the $z'_3$ pole resides on a more distant RS, 
it is the $z_3$ pole that appears relevant for any physical observables and can be assigned 
to the $\Lambda(1670)$ resonance. Nevertheless, both these poles have the same origin in the ZCL 
and both of them go to the {\bf (8b)} octet state when $x \rightarrow 0$, the $z'_3$ pole manages 
to get exactly there while the track of the $z_3$ pole gets only close to the {\bf (8b)} pole position 
due to ending up on a different (not fully symmetric) RS.

Let us continue with discussing the ratios of the channel couplings introduced in Eqs.~(\ref{eq:r1}-\ref{eq:r27}). 
Table \ref{tab:rates} provides the ratios for all four 
discussed poles and three SU(3) scaling parameter values, $x = 0,\; 0.5$ and $1$. 
When one checks the rates for $x = 0$ and compares them with the predictions made 
in the previous section we confirm our preliminary assignment of the SU(3) flavor singlet and octet 
states to the pole positions in this limit. In particular, it is the $z'_3$ pole trajectory that 
clearly goes to the SU(3) octet state while the $z_3(x = 0)$ rates do not comply with 
any of the predictions given in Eqs.~(\ref{eq:r1}-\ref{eq:r27}). Referring to the parameters 
entering Eq.~(\ref{eq:r8}), we find from the ratios $r_{12},\,r_{24}$ at $x = 0$ that 
$F_{Y}(z_{2})/D_{Y}(z_{2})\approx -1.439$ and $F_{Y}(z_{3}')/D_{Y}(z_{3}')\approx 0.357$, 
the product of the latter two ratios being $\approx -0.514$ (for which one would expect $-5/9$ 
from the decomposition of the two ortho\-gonal octet states, if the two states were exactly mass-degenerate). 

%----------------------------------------------------------------------------------------------
\begin{table}[htb]
\caption{The couplings ratios $r_{ij}$ for $x = 0,\; 0.5$ and $1$.}

\begin{center}
\begin{tabular}{c|c|ccc||c|c|ccc}
\toprule \toprule
  pole  & $x$ & $r_{12}$ & $r_{13}$ & $r_{24}$ & pole & $x$ & $r_{12}$ & $r_{13}$ & $r_{24}$ \\
 \midrule \midrule
        &   0.0     & 1.225 & 1.732 & 1.000 &        &   0.0     & 0.739 & 1.732 & 0.624  \\
 $z_1$  &   0.5     & 0.985 & 2.768 & 6.861 & $z_2$  &   0.5     & 0.826 & 1.559 & 3.805  \\
        &   1.0     & 1.380 & 5.542 & 19.04 &        &   1.0     & 0.659 & 1.635 & 17.53  \\ 
\midrule
        &   0.0     & 3.417 & 3.120 & 4.154 &        &   0.0     & 1.182 & 1.732 & 28.63  \\
 $z_3$  &   0.5     & 0.730 & 0.507 & 0.202 & $z'_3$ &   0.5     & 1.538 & 0.576 & 0.175  \\
        &   1.0     & 0.974 & 0.393 & 0.156 &        &   1.0     & 1.208 & 0.440 & 0.152  \\ 
\bottomrule
\end{tabular}
\end{center}
\label{tab:rates}
\end{table}
%----------------------------------------------------------------------------------------------

\begin{figure}[htb!]    
\centering
\includegraphics[width=0.48\textwidth]{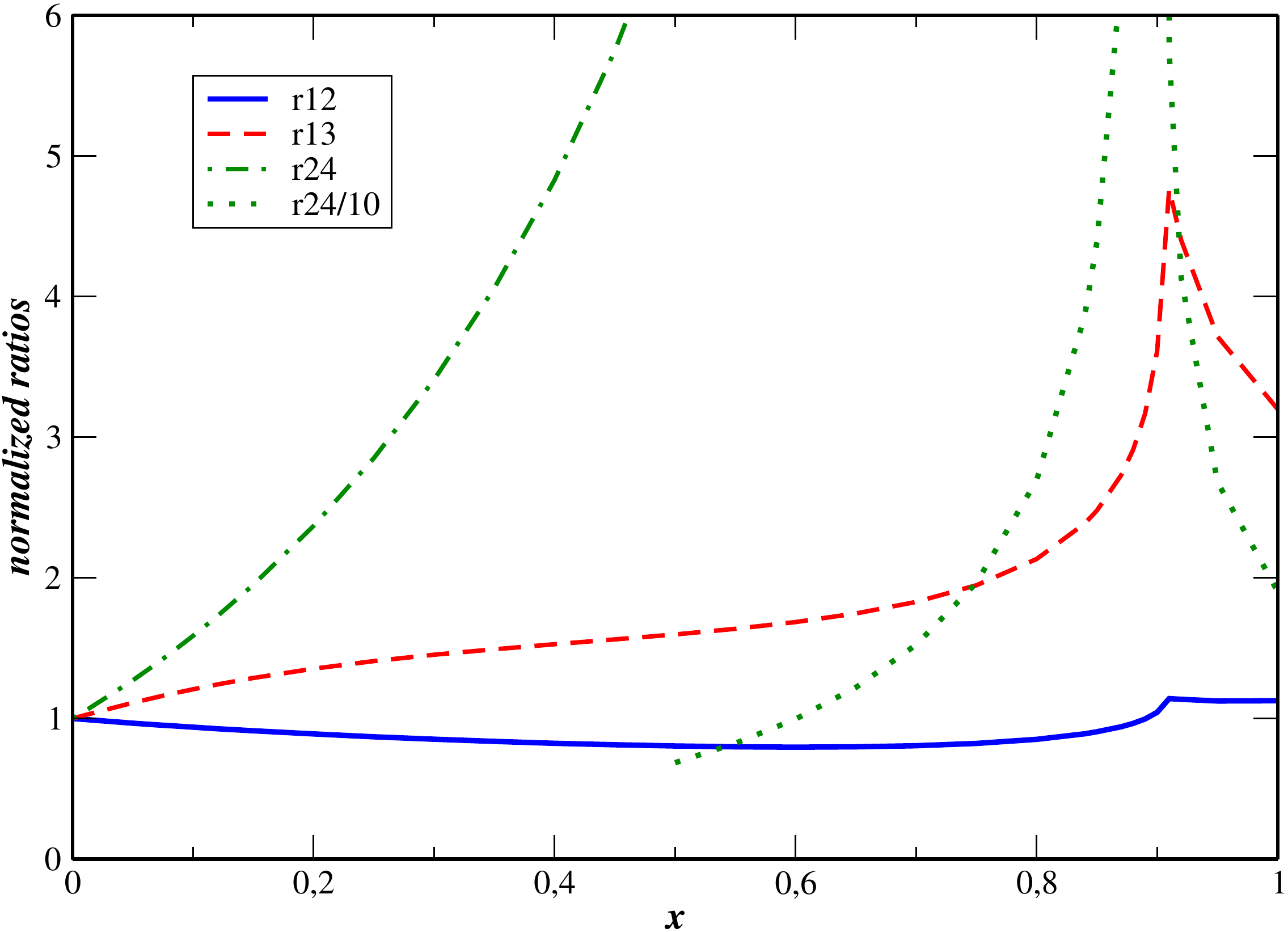} \hspace*{2mm}
\includegraphics[width=0.48\textwidth]{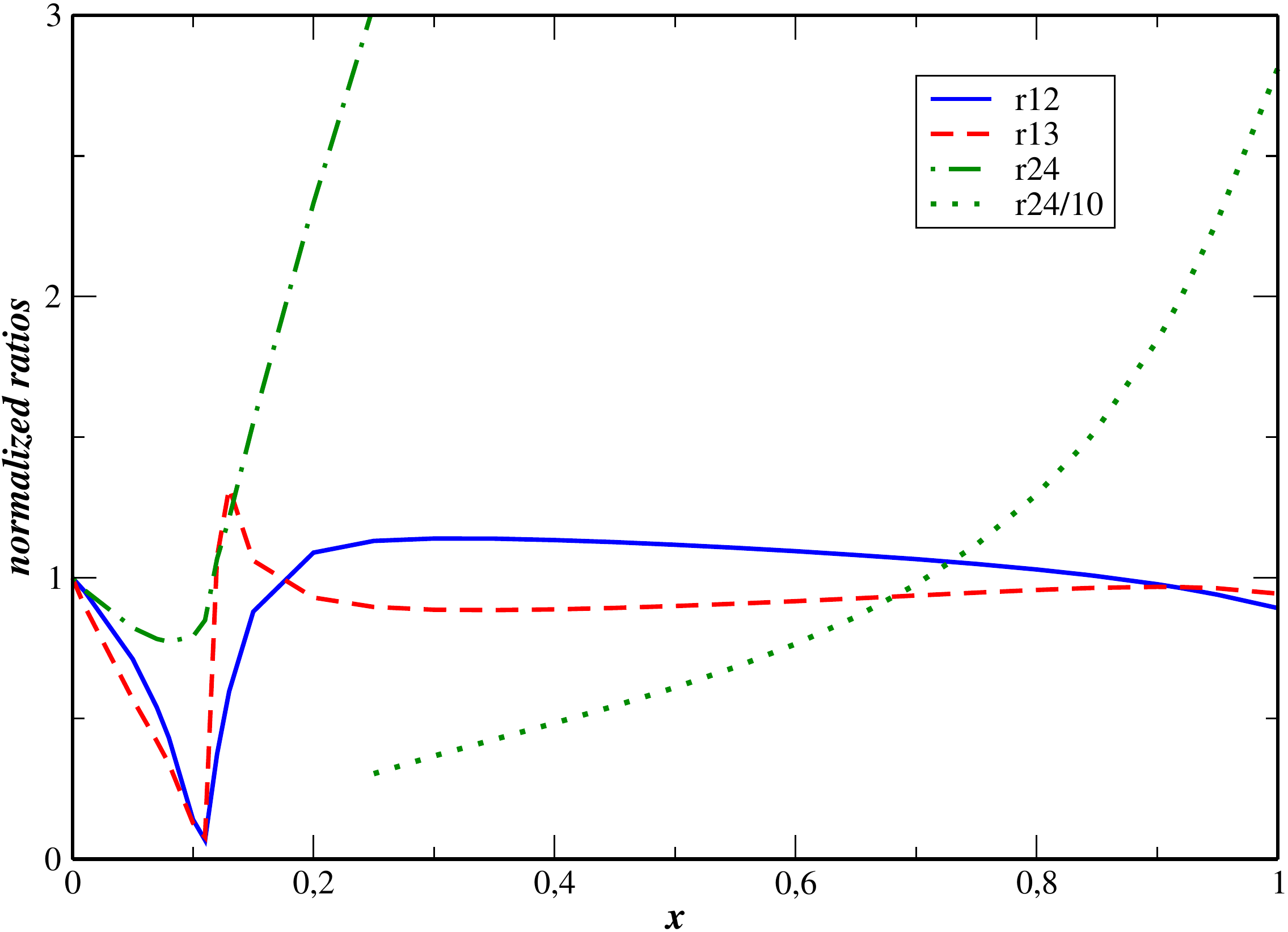} 
\caption{Dependence of the ratios $r_{12}$, $r_{13}$ and $r_{24}$ on the $x$ factor 
used to scale the breaking of the SU(3) flavor symmetry. The ratios are normalized to their 
values at $x = 0$. Left panel - the $z_1$ pole, right panel - $z_2$ pole.
}
\label{fig:ratesz12}
\end{figure}

In the physical limit ($x = 1$) our results are in agreement with observations made in many 
previous papers, see e.g.~\cite{Guo:2012vv,Feijoo:2018den,Cieply:2016jby,Mai:2014xna}. 
The $z_1$ pole couples most strongly to the $\pi\Sigma$ channel, though its coupling 
to the $\bar{K}N$ channel is also large. The couplings to the $z_2$ pole are dominated by the $\bar{K}N$ 
channel and the couplings to the $z_3$ (as well as the $z'_3$) pole by the $K\Xi$ channel. Obviously,
for all three poles the largest (and most significant) is the coupling to the channel from which 
the pole originates in the ZCL \cite{Cieply:2016jby}. However, as the latter statement is true 
in the physical limit the same cannot be said in the SU(3) flavor limit. There, all four $I=0$ 
channels couple strongly to the $z_1$ and $z_2$ poles that turn into the SU(3) singlet and octet 
states for $x \rightarrow 0$, respectively. On the other hand, the $z_3$ pole is dominated 
by couplings to the $\pi \Sigma$ channel, and the $K\Xi$ coupling is marginalized for small 
$x$ values. For both the $z_3$ and $z'_3$ poles, it is remarkable how the coupling 
of the $K\Xi$ channel changes from a dominant one in the physical limit to a marginalized one 
in the limit of restored SU(3) flavor symmetry. In quite an opposite manner, the couplings 
of the same $K\Xi$ channel to the $z_1$ and $z_2$ poles change dramatically from negligible ones 
to being comparable with those of other channels when one goes from the physical to the SU(3) 
symmetric limit. For these two $\Lambda(1405)$ poles, in Fig.~\ref{fig:ratesz12} we demonstrate the whole 
dependence of the $r_{ij}$ ratios on the scaling parameter $x$. There, it can be seen 
that while the $r_{12}$ value (equal to the ratio of $\pi\Sigma$ and $\bar{K}N$ couplings) 
remains reasonably stable, the $r_{24}$ value (ratio of $K\Xi$ and $\bar{K}N$ couplings) 
changes dramatically. We point out that such behavior cannot be explained by considering 
any associated phase-space factors that might impact on the $r_{ij}$ ratios.

%..............................................................................................
%..............................................................................................

To complement our analysis of the poles and the channel couplings we present our predictions 
for the compositeness of the dynamically generated states, see Eq.~(\ref{eq:composit}), for both 
the $x=0$ and $x=1$ limits. Our results are shown in Tables \ref{tab:comp-x1} and 
\ref{tab:comp-x0}. For the $x=0$ case, the reader can easily check that the simple rules 
$|\mathcal{C}_{\pi\Sigma}/\mathcal{C}_{\bar{K}N}|=3/2$ and $|\mathcal{C}_{K\Xi}/\mathcal{C}_{\bar{K}N}|=1$ 
apply for the singlet state, and $|\mathcal{C}_{\pi\Sigma}/\mathcal{C}_{\eta\Lambda}|=3$ for both 
singlet and octet states, compare Eqs.~(\ref{eq:r1}), (\ref{eq:r8}) and (\ref{eq:composit}). Turning 
to the $x=1$ case, it is seen that the aforementioned ratios deviate strongly from these SU(3) 
predictions. While we refrain here from joining the debate on the (probabilistic) interpretation 
of the $\mathcal{C}_{jb}$ values (see e.g.~\cite{Sekihara:2014kya,Nagahiro:2014mba,Guo:2015daa,Bruns:2019xgo}), we just point out 
that the state related to the $\Lambda_2(1405)$ pole $z_{2}$ is apparently dominated 
by a $\bar{K}N$ molecular component ($|\mathcal{C}_{\bar{K}N}(z_{2})|\approx 0.97$), whereas the state 
related to $z_{1}$ is dominated by a $\pi\Sigma$ component ($|\mathcal{C}_{\pi\Sigma}(z_{1})|\approx 0.65$). 
The higher-lying channels seem to play only a minor role in the composition of these states, 
but take a somewhat bigger part in the formation of the states related to $z_{3}$ and $z_{3}'$. 
These findings are in good qualitative agreement with earlier studies, compare e.g.~Table~1 
in \cite{Sekihara:2014kya}, Table~1 in \cite{Molina:2015uqp} and Table~4 in \cite{Miyahara:2018onh}.

%----------------------------------------------------------------------------------------------
\begin{table}[htb!]
\caption{Pole couplings and compositeness in the SU(3) flavor limit, for $x=0$. }
\begin{center}
\begin{tabular}{cc|cccc}
\toprule \toprule
\multicolumn{2}{c|}{pole} & $z_1$ & $z_2$ & $z_3$ & $z'_3$ \\
\multicolumn{2}{c|}{$\sqrt{s}$ [MeV]} & $(1482, 0)$ & $(1536, 0)$ & $(1551, 0)$ & $(1561, 0)$ \\
\multicolumn{2}{c|}{RS} & $[+,+,+,+]$ & $[+,+,+,+]$ & $[+,-,-,+]$ & $[+,+,+,+]$ \\
 \midrule \midrule
               & $\mid \beta_{jb} \mid$ & 0.479 & 0.226 & 0.629 & 0.222 \\
 $\pi\Sigma$   & $\mathcal{C}_{jb}$   & ( 0.205, 0.000) & ( 0.099, 0.) & (1.244,-0.000) & (0.484,-0.000) \\
               & $\mid \mathcal{C}_{jb} \mid$             & 0.205 & 0.099 & 1.244 & 0.484 \\ 
\midrule
               & $\mid \beta_{jb} \mid$ & 0.391 & 0.306 & 0.184 & 0.188 \\
 $\bar{K}N$    & $\mathcal{C}_{jb}$   & (0.136, 0.000) & ( 0.181, 0.000) & (-0.134, 0.) & ( 0.347, 0.000) \\
               & $\mid \mathcal{C}_{jb} \mid$             & 0.136 & 0.181 & 0.134 & 0.347 \\ 
\midrule
               & $\mid \beta_{jb} \mid$ & 0.277 & 0.130 & 0.202 & 0.128 \\
 $\eta\Lambda$ & $\mathcal{C}_{jb}$   & (0.068, 0.000) & ( 0.033, 0.000) & (-0.160, 0.000) & ( 0.161, 0.000) \\
               & $\mid \mathcal{C}_{jb} \mid$             & 0.068 & 0.033 & 0.160 & 0.161 \\ 
\midrule
               & $\mid \beta_{jb} \mid$ & 0.391 & 0.490 & 0.044 & 0.007 \\
 $K\Xi$        & $\mathcal{C}_{jb}$   & ( 0.136, 0.000) & ( 0.464, 0.000) & ( 0.006, 0.000) & ( 0.000, 0.000) \\
               & $\mid \mathcal{C}_{jb} \mid$             & 0.136 & 0.464 & 0.006 & 0.000 \\ 
\bottomrule
\end{tabular}
\end{center}
\label{tab:comp-x1}
\end{table}
%----------------------------------------------------------------------------------------------

%----------------------------------------------------------------------------------------------
\begin{table}[htb!]
\caption{Pole couplings and compositeness in the physical limit, for $x=1$.}

\begin{center}
\begin{tabular}{cc|cccc}
\toprule \toprule
\multicolumn{2}{c|}{pole} & $z_1$ & $z_2$ & $z_3$ & $z'_3$ \\
\multicolumn{2}{c|}{$\sqrt{s}$ [MeV]} & $(1353,-43)$ & $(1429,-24)$ & $(1677,-14)$ & $(1676,5)$ \\
\multicolumn{2}{c|}{RS} & $[-,+,+,+]$ & $[-,+,+,+]$ & $[-,-,-,+]$ & $[+,-,-,+]$ \\
 \midrule \midrule
               & $\mid \beta_{jb} \mid$ & 0.670 & 0.433 & 0.096 & 0.118 \\
 $\pi\Sigma$   & $\mathcal{C}_{jb}$   & ( 0.323, 0.565) & ( 0.072,-0.208) & (-0.003,-0.009) & (-0.007,-0.012) \\
               & $\mid \mathcal{C}_{jb} \mid$             & 0.651 & 0.220 & 0.009 & 0.014 \\ 
\midrule
               & $\mid \beta_{jb} \mid$ & 0.486 & 0.656 & 0.099 & 0.097 \\
 $\bar{K}N$    & $\mathcal{C}_{jb}$   & (-0.101,-0.176) & ( 0.957, 0.132) & ( 0.010,-0.004) & ( 0.000, 0.010) \\
               & $\mid \mathcal{C}_{jb} \mid$             & 0.203 & 0.966 & 0.011 & 0.010 \\ 
\midrule
               & $\mid \beta_{jb} \mid$ & 0.121 & 0.265 & 0.245 & 0.267 \\
 $\eta\Lambda$ & $\mathcal{C}_{jb}$   & (-0.005,-0.001) & ( 0.028, 0.012) & (-0.024, 0.192) & (-0.019,-0.268) \\
               & $\mid \mathcal{C}_{jb} \mid$             & 0.005 & 0.030 & 0.194 & 0.269 \\ 
\midrule
               & $\mid \beta_{jb} \mid$ & 0.026 & 0.037 & 0.633 & 0.640 \\
 $K\Xi$        & $\mathcal{C}_{jb}$   & ( 0.000, 0.000) & ( 0.000, 0.000) & ( 0.276,-0.083) & ( 0.276, 0.097) \\
               & $\mid \mathcal{C}_{jb} \mid$             & 0.000 & 0.000 & 0.288 & 0.293 \\ 
\bottomrule
\end{tabular}
\end{center}
\label{tab:comp-x0}
\end{table}
%----------------------------------------------------------------------------------------------

Let us compare our findings with an earlier analysis by D.~Jido et.~al.~\cite{Jido:2003cb} 
and discuss a possible model dependence of our results. First of all, it should be noted that in 
Ref.~\cite{Jido:2003cb}  only the WT term was considered when constructing 
the effective coupled-channel interaction potentials. In this setting the two SU(3) octet states 
merge into one and both, the $z_2$ and $z_3$ poles head to the same state when the SU(3) flavor 
symmetry is gradually restored. Our results for the $\Lambda^*$ related poles confirm the observations 
made in \cite{Jido:2003cb}, the $\pi\Sigma$ related (in terms of its origin in ZCL) $\Lambda_1(1405)$ pole 
goes to the SU(3) singlet state, while the $\bar{K}N$ related $\Lambda_2(1405)$ pole and the ($K\Xi$ related) 
$\Lambda(1670)$ pole go to the octet state. However, as our MB interaction 
is not restricted to the WT term, we were able to demonstrate that the two latter 
poles go to the {\bf (8a)} and {\bf (8b)} octets, respectively. Thus, each of the three $\Lambda^*$ states goes 
to a different SU(3) flavor symmetric state. The exact positions of the {\bf (8a)} and {\bf (8b)} octet states depend 
on the diagonalized effective potentials $v_{\bm{(8a)}}$ and $v_{\bm{(8b)}}$, and on the inverse range 
$\alpha_0 = \alpha_{jb}(x=0)$. In fact, for the Prague NLO30 \cite{Cieply:2011nq} and the current model the positions 
of the poles $z_{\bm{(\mu)}}$, where $\bm{(\mu)}$ stands for the singlet {\bf (1)} or for the octets {\bf (8a)} 
and {\bf (8b)}, can be found as a solution of Eq.~(17) in \cite{Cieply:2016jby} in which the common 
inverse range $\alpha_0$ and pertinent SU(3) symmetric $\mu$-state quantities are to be used 
instead of the $n$-th channel ones. We have checked that the two octet poles {\bf (8a)} and {\bf (8b)} 
indeed merge if we gradually switch off the Born and NLO terms contributing to the potentials 
$v_{\bm{(8a)}}$ and $v_{\bm{(8b)}}$. For the parameter setting of our current model the adopted value 
$\alpha_0 = 750$ MeV is also sufficiently large to generate both {\bf (8a)} and {\bf (8b)} poles 
on the physical [+,+,+,+] RS, with the {\bf (8b)} pole energy $z_{\bm{(8b)}}=1561.1$ MeV just below 
the common MB threshold $\sqrt{s_{thr}}=m_{N}(x=0)+M_{K}(x=0)=1562.6$ MeV. If the $\alpha_0$ value was larger,
all SU(3) symmetric $\mu$-states would be more bound and the pertinent poles would be located 
at smaller energies. On the other hand, smaller values of $\alpha_0$ lead to weaker binding 
and for $\alpha_0 = 700$ MeV the {\bf (8b)} pole effectively reaches the threshold, being located marginally 
(by $0.03$ MeV) below it. For even smaller values of $\alpha_0$ the {\bf (8b)} pole moves to the unphysical 
$[-,-,-,-]$ RS and becomes a virtual SU(3) symmetric octet state.

The sensitivity of the SU(3) multiplets pole positions to the chosen $\alpha_0$ value may 
raise some concerns. The same can be said about the {\it natural value} $a_0 \approx -2$ 
adopted in \cite{Jido:2003cb} for a subtraction constant used in 3d regularization 
of the MB loop integrals in the flavor symmetric limit. In Table \ref{tab:Amu} 
we show the SU(3) flavor symmetric scattering lengths $a_{\mu}$ at the common MB 
threshold $\sqrt{s_{thr}}$. The presented
values were obtained by transforming the scattering length matrix $a_{ia,jb} = f_{ia,jb}(\sqrt{s_{thr}})$ 
into a diagonal form with the help of Eq.~(\ref{eq:Strafo}). There is a notable sensitivity 
of the value of $a_{\mu}$ to $\alpha_0$, especially for the octet components. In particular, we note 
that the extremely large value of $a_{\bm{(8b)}}$ for $\alpha_0 = 700$ MeV is due the {\bf (8b)} pole position 
being practically at the threshold. Naturally, all the scattering lengths get smaller as the pole 
positions move down (to lower energies), away from the threshold, when $\alpha_0$ increases.
In principle, it should be possible to calculate the MB scattering lengths 
at a flavor-symmetric point within Lattice QCD \cite{Torok:2009dg,Lage:2009zv,Detmold:2015qwf,Bietenholz:2011qq,Bickerton:2019nyz}. 
Such data would certainly be helpful to fix the parameter $\alpha_{0}$ in our model. Presently, 
however, we can only offer an estimated range for our predictions, as given in Table \ref{tab:Amu}.

%----------------------------------------------------------------------------------------------
\begin{table}[htb!]
\caption{The SU(3) flavor symmetric scattering lengths $a_{\bm{(\mu)}}$, calculated for $x=0$ 
at the common meson-baryon threshold energy $\sqrt{s_{thr}}=1562.6$ MeV. The results are presented 
for three choices of the common regularization scale, the inverse range parameter $\alpha_0$.}
\begin{center}
\begin{tabular}{c|cccc}
\toprule \toprule
 $\alpha_0$ [MeV] & $a_{\bm{(1)}}$ [fm] & $a_{\bm{(8a)}}$ [fm] & $a_{\bm{(8b)}}$ [fm] & $a_{\bm{(27)}}$ [fm] \\ \midrule
    700           & -1.007     & -1.912        &   -46.861     &  -0.248    \\
    750           & -0.893     & -1.539        &   -6.755      &  -0.240    \\
    800           & -0.802     & -1.288        &   -3.640      &  -0.233    \\
\bottomrule
\end{tabular}
\end{center}
\label{tab:Amu}
\end{table}
%----------------------------------------------------------------------------------------------

We have also tested the model dependence by performing the same ana\-lysis with 
the Prague NLO30 model \cite{Cieply:2011nq}. As the model setting (chiral Lagrangian treatment, 
NLO couplings) is different, so are the pole positions in both, the physical as well as in the SU(3) 
flavor symmetric limit. In particular, the {\bf (8b)} octet pole is located on the $[-,-,-,-]$ RS 
as a virtual state just below the MB threshold $\sqrt{s_{thr}}$ when we use 
the $\alpha_0 = 750$ MeV value. Otherwise, the NLO30 model is in a nice qualitative agreement 
with the current model: the $\pi\Sigma$ and $\bar{K}N$ related $\Lambda(1405)$ poles go 
to the SU(3) singlet and {\bf (8a)} octet states, respectively, and the $K\Xi$ related 
$\Lambda(1670)$ pole goes to the {\bf (8b)} octet state but ends up (for $x = 0$) on a RS 
that is not completely channel symmetric. Exactly as in our current model, it is a shadow 
$K\Xi$ related pole that originates from the {\bf (8b)} octet state.

Finally, we mention that both our current and the NLO30 Prague model generate dynamically 
three poles in the isovector sector. The existence of such poles in the physical limit was 
discussed in \cite{Cieply:2016jby} and we checked that upon restoring the SU(3) flavor symmetry 
these poles go to the same SU(3) singlet and octet states as do the $\Lambda^*$ poles found 
in the isoscalar sector. A similar observation was made in \cite{Jido:2003cb}, though 
the trajectories of only two isovector poles were followed there. Unfortunately, the isovector 
sector is not restricted much by the current experimental data on $\bar{K}N$ interactions, 
so any predictions made for it are strongly model dependent and not very reliable. 
For this reason, we refrain from discussing any dynamically generated $\Sigma^*$ states 
within the SU(3) flavor symmetry context, at least for the time being. The situation may 
change with appearance of new experimental data in the near future, e.g.~those from measurements 
of kaonic deuterium data \cite{Curceanu:2020vjj} or on the $K^{0}_{L}$ reactions \cite{KLF:2020gai}.

\newpage
%%%%%%%%%%%%%%%%%%%%%%%%%%%%%%%%%%%%%%%%%%%%%%%%%%%%%%%%%%%%%%%%%%%%%%%%%%%%%%%%%%%%%%%%%%%%%%%
\section{Summary and conclusions}
\label{sec:summary}
%%%%%%%%%%%%%%%%%%%%%%%%%%%%%%%%%%%%%%%%%%%%%%%%%%%%%%%%%%%%%%%%%%%%%%%%%%%%%%%%%%%%%%%%%%%%%%%

In this contribution, we have presented an updated version of the model detailed in \cite{Cieply:2011nq} 
and \cite{Cieply:2009ea}, modified in the spirit of our recent work on $\eta N$ and $\eta'N$ 
interactions \cite{Bruns:2019fwi}. As we have demonstrated, we arrived at a satisfying description 
of the $I=0$, $S=-1$ MB s-wave scattering processes in the $\Lambda(1405)$ as well as 
in the $\Lambda(1670)$ resonance regions. This achievement enabled us to analyze the variations 
of the pole positions and couplings to the $\Lambda^{\ast}$ states when the flavor-symmetry breaking 
terms in our model are continuously switched off. It is worth pointing out that we necessarily need 
a good control over the amplitude in the $\Lambda(1670)$ region to study this variation with some confidence, 
because the aforementioned resonances move close together when approaching the flavor-symmetry limit 
(see our Fig.~\ref{fig:poles} and compare also Fig.~1 in \cite{Jido:2003cb}), forming one dense cluster 
of near-threshold resonances that is disentangled only by turning on the symmetry breaking. 
It is reassuring to see that the flavor classification of the resonances agrees with the earlier 
analysis performed in \cite{Jido:2003cb} employing only the WT interaction kernel and 
a somewhat different construction of the coupled-channel amplitude. Our use of a higher-order kernel 
allows us not only to confirm these findings, but also facilitates a more detailed resolution 
of the pole structure of the amplitude, applicable to a wider energy range. 

Our observations, described in the previous section, also warn us of an imprudent application 
of SU(3)-based estimates to resonances in a multi-channel scattering problem: As the couplings 
of the channels (and their ratios) to the dynamically generated resonances can vary quite 
dramatically when going from the physical situation to the limit of restored SU(3) flavor symmetry, 
the couplings determined from physical observables cannot be used immediately to reliably identify 
the flavor multiplet to which the resonance in question is (predominantly) associated. At least 
in some complicated cases, it seems necessary to follow the whole path of the pertinent pole upon 
gradually restoring the flavor symmetry to establish the pole origin in a proper way. In this respect, 
our findings put under question any such assignments made purely on the basis of partial wave analyses 
of the experimental data, e.g.~in the one performed recently in \cite{Anisovich:2020lec}. 
In particular, we have demonstrated the possibility that the trajectory of some resonance pole ends up 
(for $x=0$) on a sheet labeled with a mixed signature, to which the usual SU(3) estimates do not even 
apply in the fully symmetric situation. Though, one could probably develop generalized symmetry relations 
for such cases. To our knowledge, such a possibility is not widely discussed in the literature 
on applications of SU(3) symmetry to hadron physics. Of course, we must mention here that, 
while we have obtained a good description of the scattering data, our amplitude is not well constrained 
near the unphysical point $x=0$. There, our results should be seen as a tentative prediction. 
In principle, this situation could be improved by an evaluation of the $S=-1$ baryon spectrum around 
a flavor-symmetric point within Lattice QCD, following e.g.~the strategy presented in \cite{Bietenholz:2011qq,Bickerton:2019nyz}. 
Up to now, however, we have to rely on the emergence of a consistent picture from comparisons 
with alternative models for $\bar{K}N$ scattering. Further comparative studies in this direction 
are left for future publications.

%%%%%%%%%%%%%%%%%%%%%%%%%%%%%%%%%%%%%%%%%%%%%%%%%%%%%%%%%%%%%%%%%%%%%%%%%%%%%%%%%%%%%%%%%%%%%%%
\section*{Acknowledgement}
%%%%%%%%%%%%%%%%%%%%%%%%%%%%%%%%%%%%%%%%%%%%%%%%%%%%%%%%%%%%%%%%%%%%%%%%%%%%%%%%%%%%%%%%%%%%%%%

This work was supported by the Czech Science Foundation GACR grant 19-19640S.
%
%\newpage
%%%%%%%%%%%%%%%%%%%%%%%%%%%%%%%%%%%%%%%%%%%%%%%%%%%%%%%%%%%%%%%%%%%%%%%%%%%%%%%%%%%%%%%%%%%%%%%
\appendix
\section{Trajectory from the symmetric limit to the physical point}
\label{app:traj}
\def\theequation{\Alph{section}.\arabic{equation}}
\setcounter{equation}{0}
%%%%%%%%%%%%%%%%%%%%%%%%%%%%%%%%%%%%%%%%%%%%%%%%%%%%%%%%%%%%%%%%%%%%%%%%%%%%%%%%%%%%%%%%%%%%%%%

The flavor-symmetric limit we consider in this work is specified by the formulae below. 
%We choose the trajectory to the physical point (described by varying the parameter $x=0\ldots1$) 
%such that the average quark mass $\frac{1}{3}\left(m_{u}+m_{d}+m_{s}\right)$ stays (approximately) constant.
The trajectory to the physical point is defined by varying the parameter $x=0\ldots1$ in a way 
that the average quark mass $\frac{1}{3}\left(m_{u}+m_{d}+m_{s}\right)$ stays (approximately) constant.
We can achieve this by using the following prescriptions:
\begin{eqnarray*}
  M_{\pi}^2(x) &=& \frac{1}{3}\left(2M_{K}^2+M_{\pi}^2\right) - \frac{2x}{3}\left(M_{K}^2-M_{\pi}^2\right)\,,\\
  M_{K}^2(x) &=& \frac{1}{3}\left(2M_{K}^2+M_{\pi}^2\right) + \frac{x}{3}\left(M_{K}^2-M_{\pi}^2\right)\,,\\
  M_{\eta}^2(x) &=& \frac{1}{3}\left(2M_{K}^2+M_{\pi}^2\right) + \frac{x}{3}\left(3M_{\eta}^2-\left(2M_{K}^2+M_{\pi}^2\right)\right)\,,\\
  m_{N}(x) &=& \frac{1}{3}\left(m_{N}+m_{\Sigma}+m_{\Xi}\right) - \frac{x}{3}\left(m_{\Sigma}+m_{\Xi}-2m_{N}\right)\,,\\
  m_{\Xi}(x) &=& \frac{1}{3}\left(m_{N}+m_{\Sigma}+m_{\Xi}\right) + \frac{x}{3}\left(2m_{\Xi}-m_{N}-m_{\Sigma}\right)\,,\\
  m_{\Sigma}(x) &=& \frac{1}{3}\left(m_{N}+m_{\Sigma}+m_{\Xi}\right) + \frac{x}{3}\left(2m_{\Sigma}-m_{N}-m_{\Xi}\right)\,,\\
  m_{\Lambda}(x) &=& \frac{1}{3}\left(m_{N}+m_{\Sigma}+m_{\Xi}\right) - \frac{x}{3}\left(m_{\Sigma}+m_{N}+m_{\Xi}-3m_{\Lambda}\right)\,.
\end{eqnarray*}
Masses without argument $x$ denote the physical masses at $x=1$. On this trajectory, the quantities
\begin{displaymath}
  X_{\pi}^2:=\frac{1}{3}\left(2M_{K}^2(x)+M_{\pi}^2(x)\right)\quad\mathrm{and}\quad X_{N}:=\frac{1}{3}\left(m_{N}(x)+m_{\Sigma}(x)+m_{\Xi}(x)\right)
\end{displaymath}
stay constant, which corresponds (at next-to-leading order in BChPT) to a constant average quark mass. 
Moreover, the variation of $m_{\Sigma}(x)+m_{\Lambda}(x)$ is proportional to the famous Gell-Mann-Okubo difference,
\begin{equation*}
\Delta_{\mathrm{GMO}}:= \frac{1}{4}\left(3m_{\Lambda}+m_{\Sigma}-2m_{N}-2m_{\Xi}\right)\approx 6\,\mathrm{MeV}\,,
\end{equation*}
and thus rather small, again in agreement with NLO BChPT. It has been nicely demonstrated in lattice 
QCD simulations that flavor-singlet related quantities like $X_{\pi}^2$ and $X_{N}$ are remarkably constant 
when the flavor breaking is tuned while keeping the averaged quark mass fixed \cite{Bietenholz:2011qq,Bickerton:2019nyz}.
At the same time, the flavor-breaking is described very well by terms linear in the quark-mass difference 
$m_{s}-\frac{1}{2}(m_{u}+m_{d})$. We can thus be confident that our mass trajectories are in qualitative agreement with QCD.

In a similar manner, the meson decay constants should consistently be varied as
\begin{eqnarray*}
  F_{\pi}(x) &=& \frac{1}{3}\left(F_{\pi}+2F_{K}\right) -\frac{2x}{3}\left(F_{K}-F_{\pi}\right)\,,\\
  F_{K}(x) &=& \frac{1}{3}\left(F_{\pi}+2F_{K}\right) +\frac{x}{3}\left(F_{K}-F_{\pi}\right)\,,\\
  F_{\eta}(x) &=& \frac{1}{3}\left(F_{\pi}+2F_{K}\right) +\frac{x}{3}\left(3F_{\eta}-F_{\pi}-2F_{K}\right)\,,
\end{eqnarray*}
compare Eqs.~(16)-(20) in \cite{Bruns:2012eh}\,. Finally, the inverse ranges are also varied linearly in $x$,
\begin{equation*}
  \alpha_{jb}(x) = \alpha_{0} + x\, (\alpha_{jb}-\alpha_{0})\,.  
\end{equation*}
As we have no data to determine $\alpha_0$, the common inverse range in the symmetric limit, we simply 
adopt a reasonable value close to the average of the fitted physical inverse ranges, $\alpha_{0}=750$ MeV, 
in analogy to the treatment of the subtraction constants in Sec.~3 and Appendix~A of \cite{Jido:2003cb}. 
The sensitivity of our results to the selected $\alpha_0$ value is discussed in Sec.~\ref{sec:results}.

%%%%%%%%%%%%%%%%%%%%%%%%%%%%%%%%%%%%%%%%%%%%%%%%%%%%%%%%%%%%%%%%%%%%%%%%%%%%%%%%%%%%%%%%%%%%%%%
%\bibliographystyle{elsarticle-num}
\bibliographystyle{unsrt}
\bibliography{KbarN_SU3_cite}
%%%%%%%%%%%%%%%%%%%%%%%%%%%%%%%%%%%%%%%%%%%%%%%%%%%%%%%%%%%%%%%%%%%%%%%%%%%%%%%%%%%%%%%%%%%%%%%

\end{document}